\documentclass[a4paper,11pt]{article}
 
\usepackage{chicago}
\usepackage{multirow}
\usepackage{graphicx}
\usepackage{endnotes}
\usepackage{authblk}                                           
\usepackage{amssymb}
\usepackage{bbm}                                               
\usepackage{tensor}                                            
\usepackage{amsmath}
\usepackage[usenames,dvipsnames]{color}
\usepackage{setspace}                           

\usepackage{epigraph}
\usepackage{ulem} 
\usepackage{comment}
\usepackage{hyperref}
\usepackage[utf8x]{inputenc}
\usepackage[left=1in,right=1in,top=1in,bottom=1in]{geometry}                                  

\hypersetup{
    colorlinks=true,         
    linkcolor=black,          
    citecolor=black,        
    urlcolor=black             
}

\let\oldmarginpar\marginpar
\renewcommand\marginpar[1]{\oldmarginpar{\color{red}\raggedright\scriptsize #1}}
\newcommand{\pb}[2]{\ensuremath{\lf\{#1,#2 \rt\}}}

\def\lf {\ensuremath{\left}}
\def\rt {\ensuremath{\right}}

\title{{\bf Time Remains}}

\author[1]{ \bf Sean Gryb\thanks{email: \href{mailto:s.gryb@hef.ru.nl}{s.gryb@hef.ru.nl}}}
\author[2]{\bf Karim Th\'ebault\thanks{email: \href{mailto:karim.thebault@gmail.com}{karim.thebault@gmail.com}}}

\affil[1]{\small{{\it  Institute for Mathematics, Astrophysics and Particle Physics}, Radboud University, Huygens Building, Heyendaalseweg 135, 6525 AJ Nijmegen, The Netherlands}}
\affil[2]{\small{{\it Munich Center for Mathematical Philosophy}, Ludwig Maximilians Universit\"{a}t, Ludwigstrasse 31, D-80539, Munich, Germany} }  

\begin{document}

\maketitle

\begin{abstract}      
On one popular view, the general covariance of gravity implies that change is relational in a strong sense, such that all it is for a physical degree of freedom to change is for it to vary with regard to a second physical degree of freedom. At a quantum level, this view of \textit{change as relative variation} leads to a \textit{fundamentally timeless} formalism for quantum gravity. Here, we will show how one may avoid this acute `problem of time'. Under our view, \textit{duration is still regarded as relative, but temporal succession is taken to be absolute}. Following our approach, which is presented in more formal terms in \cite{gryb:2014}, it is possible to conceive of a genuinely dynamical theory of quantum gravity within which time, in a substantive sense, remains.
\end{abstract}

\tableofcontents
 \newpage

 \setlength{\epigraphwidth}{.6\textwidth}
\epigraph{\textit{We should not be able to tell the story of our relations with another, however little we knew him, without registering successive movements in our own life. Thus every individual -- and I myself am one of those individuals -- measured duration by the revolution he had accomplished not only round himself but round others and notably by the positions
he had successively occupied with relation to myself.}}{\textit{\\Time Regained} \cite{proust:1931}}

\section{Introduction}

\subsection{The Problem of Time}

A key feature of Einstein's theory of gravity is its invariance under arbitrary transformations of the spacetime manifold. This \emph{diffeomorphism} symmetry implies that only the coordinate-free information contained in the geometry has a physical basis within the theory. Unfortunately, it is not entirely clear how one should understand the implications of diffeomorphism invariance for the specific role of time in the theory. In the Lagrangian formulation, where the theory is expressed in terms of the Einstein-Hilbert action, this difficulty manifests itself in our inability to find a representation of time in terms of an action of the real numbers implementing time translations on the space of physical (i.e., diffeomorphism invariant) solutions.\footnote{See \cite{Belot:2007} for an extensive discussion of this and related points regarding the representation of time and change within Lagrangian field theories, including general relativity. With regard to the Hamiltonian framework, the present analysis differs on several key interpretational and formal points. See \cite{thebault:2012b} and \cite{pitts:2014} for critical discussion.} Similarly, in the Hamiltonian formulation of the theory, which is the basis for many modern approaches to the quantization of gravity, we find ourselves lacking a coordinate free means of representing time.\footnote{This is one key aspect of the `problem of time' in quantum gravity. See \cite{Isham:pot_review} for a classic or \cite{Anderson:PoTReview} for an updated review.}

Imagine a loaf of bread that we can irregularly cut up into a sequence of slices. The loaf is spacetime and the slices are instantaneous spatial surfaces. A \textit{foliation} is then a parameterization of a spacetime by a time ordered sequence of spatial slices.
Such a parametrization is local in the sense that it is defined for every point on every spatial slice.   \emph{Diffeomorphism invariance} implies that spacetimes described by general relativity that are related by \textit{re}-foliations are physically equivalent. Within the Hamiltonian formulation, which dates back to \cite{Dirac:1958b}, we make the restriction to spacetimes that admit a foliation into such sequences of space-like hypersurfaces (the globally hyperbolic spacetimes \cite{Geroch:1970}). Spacetime diffeomorphism invariance is implemented in two parts: i) spatial diffeomorphism invariance; and ii) spacetime foliation invariance.\footnote{Note that this does not constitute the full group of spacetime diffeomorphisms since neither large diffeomorphisms nor diffeomorphisms that fail to preserve the space-like embedding of hypersurfaces are represented in the Hamiltonian formalism.} We thus  have within the theory an ability to re-slice a spacetime into an infinite number of different decompositions of space and time without changing anything physical. It is the conceptual and technical complications involved in representing this symmetry that leads to the acute `problem of time' within the formalism.

Foliation symmetry further implies that any observable quantity within the theory must not be dependent upon the local temporal labelling of spacetime.\footnote{Formally, this is usually expressed in terms of the requirement that the functions that represent observable quantities should commute with the Hamiltonian constraints, which are taken to implement foliation invariance \cite{Bergmann:1961}. There are, however, many subtleties regarding both the role of Hamiltonian constraints and the definition of observables in canonical general relativity  \cite{anderson:2013,pitts:2014}.} This leads us directly to the question of how we should understand the \textit{change} in physical quantities? In addition to not having a representation of time, we seem also to have lost a clear methodology for representing change! Our conceptual machinery appears in need of retooling.   
 
According to the \textit{correlation} or \emph{partial observables} view of time in general relativity, the radical moral one should draw from diffeomorphism invariance is that \textit{change is relational} in a strong sense, such that all that it is for a physical degree of freedom to change is for it to vary with respect to a second physical degree of freedom; and there is no sense in which this variation can be described in absolute, non-relative terms.\footnote{The correlation view is most closely associated with the work of Carlo Rovelli \cite{Rovelli:2002,Rovelli:2004} and is put forward in slightly different ways by many physicists working on canonical quantum gravity, for example see \cite{Dittrich:2006,Dittrich:2007,Thiemann:2007,bojowald:2011}. For a more detailed appraisal of the strengths and weaknesses of this view in the classical gravity context, see \cite{thebault:2012b}. Also see \cite{Rovelli:2007} for discussion of some subtitles regarding the interpretation of the scheme.} This \textit{radical relationalist} view of time implies that there is no unique parameterization of the time slices within a spacetime, and also that there is no unique temporal ordering of states. Furthermore, it implies a fundamentally different view of what a degree of freedom actually is: such parameters no longer have distinct physical significance since they can no longer be understood as being free to change and be measured independently of any other degree of freedom. This means that all one-dimensional systems must be understood as stationary since a relational notion of change cannot be constituted: there is no degree of freedom for the system to change with respect to. A one-dimensional pendulum is thus, under this understanding of dynamics, a stationary system with no genuine degrees freedom. And a two-dimensional pendulum is to be understood as a one-dimensional system, with the change in the (arbitrarily chosen) free variable expressed in terms of the other `clock' variable. 

On this view, it should be no great surprise  that when the equations of a classically foliation-invariant theory are quantized, one arrives at a timeless quantum gravity formalism\footnote{This is the Wheeler--DeWitt-type `frozen formalism', endemic within both the old quantum geometrodynamics approach \cite{DeWitt:1967}, and modern variants of canonical quantum gravity \cite{Thiemann:2007}.} -- since, in essence, this facet is already implicit within the classical theory. Both classically and quantum mechanically, the functions  that faithfully parametrize the true degrees of freedom of the theory -- the observables -- are taken to be those which are completely independent of the local time parametrization and, both classically and quantum mechanically, these \textit{perennials} cannot, by definition, vary along a dynamical trajectory. Thus, we see that this first response to the problem of time in classical and quantum gravity is essentially one of capitulation. The definition of an observable quantity within the correlation view is such that it cannot change along a dynamical trajectory. Although, we can  recover a weak sense of change as relative variation, there is no scope for the basic one-dimensional ordering structure that, in our view, is constitutive of time. To us this seems unsatisfactory as a solution, and in the remains of this paper we will articulate an alternative.

\subsection{Our Solution}    

The starting point of our approach is the conviction that the radical variant of relationalism with regard to change and time discussed above has gone a little too far. The lessons for time drawn from general covariance are more subtle, do not imply we should dispense with time altogether. Rather, a consistent interpretation of the \textit{underdetermination of possibilities} implied by the temporal relabelling symmetry of the theory, leads to a view in which \textit{temporal succession} is understood as absolute. Temporal relabelling symmetries do not result in identification of instantaneous states as identical physical possibilities and thus the formalism of theories with these symmetries can be interpreted such that both the change in a given  degree of freedom and the ordering of such change along a dynamical history are fundamental structures in the theory.\footnote{It is important to note here that by `time ordering' we specifically \textit{do not} mean anything like an arrow of time, or objective difference between past and future. Rather, `time ordering' here, and in what follows, merely implies the existence of a monotonically increasing parameterization of time slices which is, by definition, time-reversal invariant.} Given that, one can exploit the interpretational underdetermination present to take a further step regarding the status of duration. One may insist that although change itself is taken to absolute, the \textit{labelling of change} in terms of a measure of duration, is something purely relative. Such a \textit{Machian} view of time is simple both to motivate and to realise within classical particle models with global temporal re-labelling symmetry, but far less easily constituted within the full general theory of relativity due to the need to define observables that respect foliation invariance. It is therefore understandable that the radical morals with regard to change and time discussed above are often drawn. \textit{However}, it is still true that within general relativity there exist fundamental temporal structure relating to ordering in time. In canonical general relativity, such structure is encoded within the fact that the arbitrary slicings are always labelled in terms of a \textit{monotonically increasing} local time parameter (as implied by the positivity of the lapse multiplier). This structure is also present in a more subtle sense within the Lagrangian theory due to the form of the Einstein--Hilbert action. This is because the variation of the Einstein--Hilbert action (subject to the appropriate boundary conditions)\footnote{See \cite{York:GR_boundary} for a discussion of this variational principle in the context relevant here. } requires finding a curve that minimizes the integral of the scalar curvature, and these curves, by definition, require parametrization by a  monotonically increasing parameter. Thus, the formalism of general relativity should not be seen as telling us to dispense with time ordering altogether. 

Furthermore, on our view, that the (canonical) quantization of general relativity leads to a timeless formalism should be understood as a consequence of an incorrect treatment of the temporal symmetries of the classical theory. By treating local temporal labellings as \textit{entirely} unphysical, and change as \textit{entirely} relational, we do not retain in the quantum formalism the full classical dynamics or the implicit temporal ordering structure.\footnote{Equivalently, in more formal language, if  Hamiltonian constraints are treated as generating purely unphysical transformations one does not, in the quantum theory, retain their role in generating dynamics of the quantum state, nor in providing a temporal ordering.} The question remains, however, if conventional quantization techniques cannot preserve the essential temporal structure of general relativity, how do we find a new methodology that can?

Our solution involves two non-trivial steps, the motivation for which will be outlined at length below. The first relates to a fundamental re-description of gravity in terms of the \textit{shape dynamics} formalism originally  advocated by Barbour and collaborators \cite{barbour:scale_inv_particles,Barbour:2011dn,barbour_el_al:scale_inv_gravity,barbour_el_al:physical_dof} and then brought into modern form in \cite{gryb:shape_dyn}. From the view of the current paper (see also the more detailed argument of \cite{gryb:2014}), the existence of shape dynamics as, in a precise sense, a \textit{dual} to general relativity reveals classical gravity to be essentially \textit{Janus-faced}. There exist an \textit{underdetermination of symmetries} that leave space for interpretation in terms of two distinct gravitational ontologies: the traditional `Einstein ontology' of spacetimes invariant under four dimensional coordinate transformations; and a second, hitherto masked, ontology of sequences of \textit{scale-invariant} three dimensional spatial surfaces (i.e., spatial geometries invariant under re-scallings of lengths). As shall be detailed later, this second ontology is closely related to a proposal for the interpretation of the degrees of freedom of gravity made by James York \cite{York:GR_boundary}, and so we will call it the `York Ontology'.

\bigskip

\centerline{
\includegraphics[width=0.5\textwidth]{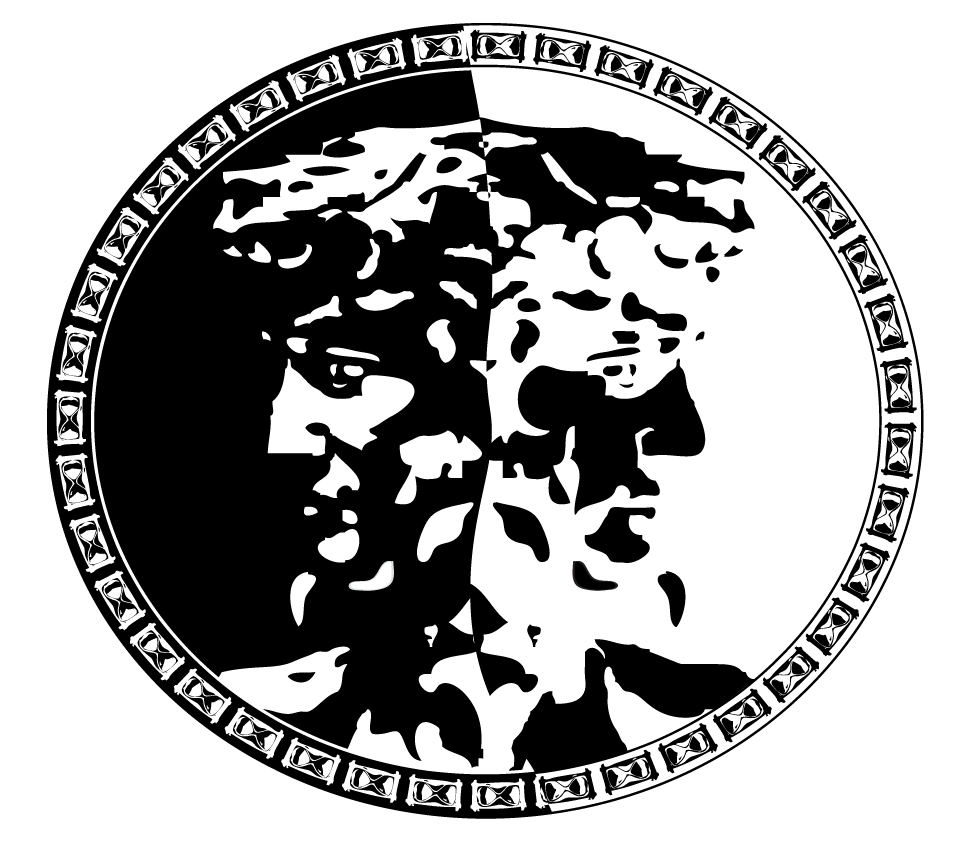} }
\begin{flushleft}
\small{Figure 1. Janus was the Roman god of gateways, transitions and time, and is usually depicted as having two, non-identical faces pointing in opposite directions as above. Thus, figuratively (and in fact geometrically) the two faces of Janus are an apt representation for the two faces of gravity.}
\end{flushleft}

Given this understanding of gravity as having two dual faces, when confronted with the problem of understanding the role of time in classical gravitation, one has the option of choosing whichever formalism -- shape dynamics or general relativity -- is formally and conceptually easier to work with. Here, we choose to use shape dynamics, and assume the particular characterization of the theory is given by the York ontology. It is from this basis that the second step in our proposal can be made. Originally in \cite{gryb:2012} and then more recently in \cite{gryb:2014}, a procedure for the \textit{relational quantization} of theories with temporal relabelling symmetries was outlined. Whereas the first of these two papers served to offer a range of conceptual arguments leading to the need for relational quantization, the second served to place relational quantization within a formal framework for understanding symmetries in physical theory in general. 

One of the principal motivations of the current paper is to explicate further the philosophical foundations of this approach to symmetries and time. In particular, in what follows we will: First, introduce a general methodology for the classification of symmetries and symmetry-related variables according to physically motivated criteria, see \S\ref{sec: Formalism}-\ref{sec: Freedom by Degrees}; Second, demonstrate that our classification scheme leads naturally to a procedure for quantization via the introduction of dummy variables, see \S\ref{sec: Voluntary Redundancy}; Third, provide philosophical motivation for the Machian view of time discussed above, see \S\ref{sec: Change and Order}; Fourth, show how our philosophical motivations for the treatment of time symmetries mesh with our general prescriptions for symmetry and lead to the procedure for relational quantization, see \S\ref{sec: Quantization and Succession}. These arguments establish a framework sufficient to motivate the relational quantization of gravity from the perspective of the York ontology, see \S\ref{sec: Two Faces} and \S\ref{sec: Retaining Succession}, and in doing so provide a demonstration that, given suitable starting assumptions, time can remain in quantum gravity.

\section{Understanding Symmetry}

\subsection{Mechanics and Representation}

\label{sec: Formalism}

In a classical physical theory the relationship between a mechanical system and its theoretical representation can be given through the specification of three pairs of structures, one of each pair formal and one of each pair physical. These pairs relate in turn to: degrees of freedom, dynamical laws, and the specifiable initial (or boundary) data. Explicitly, for a finite dimensional classical system, we can consider a physical system as being represented formally by: i) a configuration space with $n$-configuration variables and some pre-determined metric structure; ii) a \textit{nomological} restriction on curves in the configuration space that selects only curves that are geodesic with respect to the pre-determined metric structure; iii) a set of further initial (or end-point) conditions on the curves.\footnote{It can be shown, see for instance \cite{Lanczos:1970}, that these three requirements (which are collectively known as \emph{Jacobi's principle}) are equivalent to Hamilton's variational principle for mechanics. } Such a specification serves as a representation of: a) the physical degrees of freedom; b) the dynamical law which determines the evolution of the system; and c) the physical conditions on the preparation of the system. The representational pairings are then i-a, ii-b, iii-c. We will designate the formal side of these pairings the \textit{formalism} of the system and the physical side the \textit{characteristic  behaviour} of the system.  

One can be more explicit with regard to the formalism side of the set up by defining the Lagrangian formalism. For an $n$ dimensional system, the configuration space is a manifold, $\mathcal{C}$, with elements $q_i$, $i=1,...,n$. At a given point $q\in \mathcal{C}$ we can define a tangent space $T_{q}\mathcal{C}$. The disjoint union of all
the tangent spaces of $\mathcal{C}$ is the tangent bundle $T\mathcal{C}$. The elements of the tangent bundle are pairs $(q,\dot{q})$ of configuration variables $q$ and vectors tangent to
those variables $\dot{q}$. A curve within the tangent bundle, $\gamma:
\mathbb{R}\rightarrow T\mathcal{C}_{0}$, will correspond to a history of a system: a
sequence of configurations and velocities. The parameterisation of the curve is given by some monotonically increasing parameter, $t$. We then define the Lagrangian$,$ $L:T\mathcal{C}\rightarrow\mathbb{R}$, and the action, $I[\gamma ]=\int_{\gamma}L[q_{i},\dot{q}_{i}]dt$. Given the specification of a curve (including its endpoints), the extremisation of the action, $\delta I[\gamma ]=0$, according to the principle of least action leads to the Euler--Lagrange equations, $\frac{d }{dt}\left( \frac{\partial L}{\partial \dot{q}_{i}}\right) =\frac{\partial L}{\partial q_{i}}$, that specify a set of solutions, $\gamma _{S} $, which uniquely determine the possible classical histories of the system given an initial point in $T\mathcal{C}$. We now have a formal representation of physical degrees of freedom in terms of the velocity and configuration variables; the relevant nomological restrictions, in terms of the variational principle and Lagrangian; and the preparation conditions, in terms of the specification of the end points of the curves in the variation.

An alternative, but generally equivalent, \textit{Hamiltonian formalism} can then be derived by defining a cotangent bundle of our configuration
manifold: the phase space $\Gamma =T^{\ast }\mathcal{C}$. This is the disjoint union of all the cotangent spaces $T_{q}^{\ast }\mathcal{C}$ (these are dual to the tangent spaces -- i.e., elements of the cotangent space are linear functionals on the tangent space). A point in phase space, $(q,p)$, consists of a point
in our original configuration space, $q\in \mathcal{C}$, paired with a covector at $q$, $p\in T_{q}^{\ast }\mathcal{C}$. These covectors, which
we call the conjugate momenta, are given by, $p_{i}=\frac{\partial L}{\partial \dot{q}_{i}}$. We can equip phase space with a symplectic structure by defining the symplectic potential $\theta = p\wedge \text dq$ and the (closed) symplectic 2-form $\omega = \text d\theta$. If $\omega$ is non-degenerate, then one can equip phase space with a Poisson bracket such that $\pb{q^i}{p_j} = \delta^i_j$. Using this Poisson structure, we can define a \emph{Hamilton vector field}, $v_f(g)$, for any phase space functions $(f,g):\Gamma \to \mathbbm R$, via $v_f(g) = \pb g f$. We can then define the \emph{flow} of some function $f$ by the integral curves of its Hamilton vector field.

To fix the dynamics, we introduce the Hamiltonian functional, $H[q_{i},p_{i}] =p^{i}q_{i}-L$, and derive Hamilton's equations, $\dot{p}_{i}=v_H(p)$ and $\dot{q}_{i}=v_H(q)$. The classical trajectories, $\bar\gamma_S$, are, thus, uniquely defined by the flow of the Hamiltonian functional on phase space. For the purposes of this formalism, the preparation conditions are represented by specification of initial values of the positions and momenta variables.  

All this is familiar to anyone with basic knowledge of mathematical physics. It is important, however, to be clear regarding the relevant representational relationships. For a given system we have, on the one hand, the physical characteristics of the system contained in the nature of the degrees of freedom, the physical preparation of the system, and the law-like regularities in behaviour; and, on the other, we have the formal description of the system contained in the formal variables, the formal initial (or end-point) conditions, and nomological restrictions on these. The connection between these two triples is precisely what a physical theory gives us. In some very simple cases, such connections can be understood unambiguously; however, invariably, when we want to understand realistic systems, we often run into situations where the relationships between the formal and physical aspects is no longer one-to-one. In such situations, we can say that some form of \textit{representative redundancy} is present within our formal-physical set up. One of the main aims of this paper is to detail a new  scheme for understanding this representative redundancy. In particular, we aim to provide a general and physically motivated classification of different types of \textit{symmetry transformations}. This classification will entail which formal differences have the capability of representing distinct physical possibilities and which correspond to pure redundancy.

Such a symmetry classification scheme has direct implications for the \textit{ontology} that can be associated with a physical theory. At a classical level, this is because, by placing constraints on which formal differences can correspond to different physical possibilities, our understanding of symmetry also places constraints on the consistent interpretations of a theory. These constraints will invariably still leave the ontology underdetermined: the same possibility counting can be consistent with very different ontologies for a physical theory. However, in some specific circumstances -- including the case of time parametrization symmetries that is our main focus -- difference as to the symmetry classification scheme can have crucial ontological importance. The ontological importance of symmetry classification is even greater considered in the context of quantization. As we shall see in detail later, the way in which we treat symmetries classically determines which possibilities will be realised at the quantum level. 

Clearly, then, a symmetry classification scheme is something of great importance, and should be built upon a chain of sound physical and formal inferences. What is needed are general principles that are based upon physical reasoning but lead to precise mathematical diagnoses. Unfortunately, from our perspective, what is provided by most existing approaches are precise mathematical principles leading to a diagnostic schema that is neither physically well motivated nor formally rich enough. 

The standard classification of symmetries follows a scheme where the existence of local or `gauge' symmetries -- i.e. transformations that depend on space or time and under which the action is invariant -- indicates that identical physical possibilities are being represented in terms of distinct instantaneous states in the formalism.\footnote{The most comprehensive modern book on gauge theory and its quantization is \cite{Henneaux:1992a}, the original  classic is \cite{Dirac:1964}. See \cite{BelotEar:2001,Belot:2003,Earman:2003,Rickles:2004,Rickles:2007} for philosophical analysis of the connection between possibility spaces and this notion of gauge symmetry. The crucial factor in most of these accounts is the `indeterminism' that can result from not eliminating the excess possibilities associated with local symmetries. See \cite{thebault:2012a} for details on why such arguments are not decisive in general. See \cite{Pons:2005} argument that the standard Dirac analysis of canonical quantization is incomplete.} The representative redundancy inherent in such symmetries is directly connected to otiose degrees of freedom which are then eliminated during (or before) quantization. In this standard scheme, we can remain agnostic regarding the status of redundancies associated with global symmetries -- i.e.  transformations that do not depend on space or time and under which the action is invariant. The representative redundancy related to such symmetries need not be eliminated during (or before) quantization. Thus, within this scheme, the difference between local and global symmetries has huge potential impact at the level of both classical and quantum mechanical interpretation. 

In our view, the weight placed upon the local vs. global distinction represents a first major problem with the scheme. There are theories which display local symmetry but do not feature identical physical possibilities represented in terms of distinct instantaneous states. In such theories, there are in fact no excess degrees of freedom and, thus, the application of standard `local' symmetry quantization techniques will lead to the elimination of genuine physical differences. Moreover, the loss of representative machinery will have direct interpretational consequences: there will be ontologies which are excluded from the cast of possible interpretations for no good reason.

A second problem with the standard scheme is that it is ill-suited for dealing with a recently discovered form of symmetry: the class of \emph{hidden} symmetries. These symmetries are of a peculiar form, in that they are derived from implicit rather then explicit redundancies in the formal-physical relationship. When hidden symmetries occur, we are able to `trade' a certain symmetry of the theory for a second set of transformations which was originally not a symmetry. Such symmetries, thus, lead to a different form of interpretational underdetermination since we have a choice as to different formulations of a theory with different symmetries -- over and above the way we understand the implications of these symmetries. 

Each of these problems with the standard scheme is directly related to issues regarding time and gravity. With regard to the first problem, time reparametrization symmetries are an example of local symmetries which should not be associated with excess degrees of freedom. The standard scheme leads to an erroneous categorisation of dynamically related instantaneous states as constituting the same physical possibility. This rules out an interpretation of the temporal ontology of the theory in hand in terms of anything other than radical relationalism. With regard to the second problem, the identification of hidden symmetries provides the basis of the duality between General Relativity and Shape Dynamics mentioned above. It is thus only through the understanding of hidden symmetries that the `dual faces' of gravity can be identified. We will return to these points in detail when we enter into the specific discussions of time (\S\ref{sec: Understanding Time}) and gravity (\S\ref{sec:  Time and Gravitation}). Before then, we would like to construct our own scheme for the categorisations of symmetries which does not suffer from the identified defects. In order to do this, we need to find general \textit{physical} principles for distinguishing types of redundancy.

\subsection{Freedom by Degrees} \label{sec: Freedom by Degrees}

The first physical principle for distinguishing types of redundancy we can identify relies upon the notion of action, $I[\gamma]$. As noted above, the action is directly connected to the nomological restrictions that allow the formalism of a theory to pick out physical dynamics. If we consider the infinitesimal variation of a curve in a particular direction in the configuration space (by `direction' we will mean along an infinitesimal segment of the flow of a particular phase space function) and find that the action is invariant up to a total derivative, then the degree of freedom associated with that direction (i.e., the degree of freedom identified with the particular phase space function whose flow is in the relevant direction) is related to a \textit{manifest} form of symmetry. It is important to point out, for our considerations, that a manifest symmetry can either be a local symmetry or a global symmetry. This criteria relates to a property of the action itself and has nothing to do any additional variational principles one would like to further impose on the action (for example, to extract classical equations of motion). To fully classify a degree of freedom, we will need a second criteria, described below, which leads to a richer set of physically distinct cases than what is \emph{explicitly} considered by standard textbook definitions of symmetry.

Physically speaking, a manifest symmetry implies that there will be multiple possible sequences of configurations (representing histories in the configuration space) which are either \textit{physically indistinguishable} or correspond to different values of a conserved charge. In the first case, the symmetry corresponds to a mathematical transformation within the formalism that has no effect on physically measurable quantities. Thus, it can apply to the Universe \emph{as a whole}. In the second case however, the transformation has an experimental implication: it changes the value of the conserved charge. The only way to attribute meaning to this is to have an emergent structure within the formalism that allows one to measure changes of the variable conjugate to the relevant conserved charge. Thus, the second case is only relevant to sub-systems of the Universe that are dynamically isolated from the rest of the Universe, which serves as an emergent background. For a more detailed discussion of how these backgrounds emerge, see \cite{Giulini:2013aa}. 

Furthermore, the physical indistinguishability of the histories could be due either to practical limitations within the particular experimental set-up being considered (as is nicely illustrated, for example, in \cite{Wharton:2009kp}) or to fundamental limitations within the system. On its own, the presence of a manifest symmetry \textit{does not} indicate that identical physical possibilities are being represented in terms of distinct instantaneous states in the formalism.

Given an orbit on configuration space, generated by the flow of some phase space function, in which the action is \emph{not} invariant, there are two further possibilities for the relevant degree of freedom identified with the phase space function in question: it might be the case that there are \textit{no symmetries} associated with it, or it might be the case that there are \textit{hidden symmetries} associated with it. This third case  has not been explored until recently, but will prove important for gravity. We will return to its detailed consideration later. 

The second principle we can identify relates specifically to the nature of the variational principle used to vary the relevant variables in the action. As was noted above, it is essential to remember that the abstract initial (or end-point) conditions used in the variational principle are part of a formalism which stands in a possibly non-unique representative relationship with a class of physical systems defined in terms of their characteristic behaviour (i.e., physical observables within the system, physical preparations of the system, and dynamical laws obeyed by the system). Let us again consider the variation of the action based upon the variation of a configuration space curve in a direction associated with a particular degree of freedom (as defined more precisely above). In this case, let us focus upon the details of the variational principle used along the \emph{entire} history of the system in question, including the boundary (in space and time). 

If the characteristic behaviour -- which is fixed by the actual physical degrees of freedom, the physical preparation of the system, and the dynamical law -- places no restriction upon such a variation, then we say that it is a \textit{free variation}. Under our proposed scheme, this type of variation implies that the relevant degree of freedom is an otiose formal artefact since nothing in `the world' places a restriction upon the relevant variable's value. Nothing in the characteristic behaviour of the system fixes anything in the mathematical formalism, and so it must be a facet of redundancy within our representation. The alternative is that, for a given direction and associated degree of freedom, restrictions are placed on the variation. In such a situation, we say that we have a fixed endpoint variation (because the end points of the variation are fixed in configuration space), or \emph{fixed variation} for short, and we expect that the relevant degree of freedom has some representative relationship to something physical.

It is important to note why one would expect that it is the second, and not the first principle, which is decisive in the categorisation of a symmetry-related degree of freedom as inherently redundant or not. Although the first principle does derive from conditions on the action, it does not derive from conditions specifically on the variation principle. It is precisely the variational principle that fixes the full characteristic behaviour of the system. Thus, it is only the variation principle that can ultimately be sensitive to the difference between redundancies that are linked to dynamical conservation properties, and those that are entirely due to our use of excessive coordinates within the instantaneous representation of a physical state.        

Keeping this important point in mind, and given our two \textit{physically motivated} principles, we can set about categorising types of redundancy according to a \textit{physically motivated} diagnosis. One would then hope that the mathematical exactitude of standard techniques will be recoverable where these techniques have proven physically reliable. This indeed proves to be the case if we consider the most blatantly unphysical form of redundancies: those connected to manifest symmetries and free variations. In such situations, the relevant degree of freedom encodes no dynamical information, and defines a direction which is, by definition, superfluous to the representation of the world. Such variables are \textit{gauge variables} and the relevant symmetries are \textit{gauge symmetries}. Thus, we see the precise reasons why gauge symmetries (as classified by our scheme) will lead to  identical physical possibilities being represented in terms of distinct instantaneous states in the formalism.

Our categorisation is sufficient, \textit{although not necessary}, to recover the conventional mathematical categorisations of gauge symmetries \cite{Henneaux:1992a}. As discussed above, the standard account of gauge symmetries is usually in terms of a local (i.e., functionally dependent on space and time)  transformation that leaves the action (or Lagrangian) invariant. Our categorisation of gauge symmetry is sufficient (but not necessary) for this locality condition. It is also sufficient (but not necessary) to recover the other standard notions of gauge symmetry in terms of: i) the failure for the the Legendre transformation between the velocity-configuration and phase space to be invertible, and; ii) existence of first class primary constraints \cite{Dirac:1964}. This means that for all the standard cases where there is no perceived ambiguity about the cause and interpretation of redundancy -- e.g., electromagnetism, Yang-Mills theories, the standard model (all in the presence of no spatial boundaries) -- our definition will coincide with the standard definition. However, as we shall see, for the case of time labelling symmetries, and indeed many other symmetries which will not be discuss at length here,\footnote{Note that asymptotically flat GR and Yang--Mills theory in the presence of spatial boundaries which break gauge invariance are examples of theories that also \emph{do not} fall into the category of pure gauge theories by our classification, although they would by the standard treatments \cite{Henneaux:1992a}. We believe that our classification scheme is more appropriate for these cases because a notion of background is introduced by the relevant boundary conditions.} our scheme still allows for an alternative, physically motivated categorisation.

Here, we should note an important point for the purposes of our discussion: those degrees of freedom identified as gauge within the classical theory are always (in some sense) eliminated within the process of constructing the quantum theory. Since these degrees of freedom are not representing anything about the physics of the system, the quantum correlates of these degrees of freedom must not be associated with observable operators in the quantum formalism. In essence, the methodology for ensuring faithful treatment of gauge degrees of freedom is always the same: treat the direction associated with the degree of freedom as non-physical.

A further, more subtle form of redundancy derives from the presence of a degree of freedom associated with a manifest symmetry, but a fixed variation. Since the variation is fixed we know that the relevant variable \textit{is} connected to something physical. In general, we can understand this `something physical' as the conservation of some empirically determinable quantity throughout the system's evolution. For this reason, we call the relevant symmetries \textit{conservation symmetries}.\footnote{By this terminology, we do not wish to imply that \emph{all} conserved quantities are related to conservation symmetries (e.g., electric charge is a conserved quantity arising in gauge theories and is not a conservation symmetry by our definition), although conservation symmetries necessarily have conserved charges associated with them. Their distinguishing feature, which is relevant for us here, is that, in the quantum theory, conservation symmetries allow for superpositions of states with different values of the conserved charge.} There is no necessary connection between the existence of conservation symmetries and identical physical possibilities being represented in terms of distinct instantaneous states (or sequences of states) in the formalism. It is a further interpretational step, needing further motivation, to eliminate such potential redundancy. We can thus understand the interpretational implication of the existence of a conservation symmetry in terms of an \textit{underdetermination of possibilities}. One has the freedom to classify the transformations related by the symmetry as connecting the same or different physical possibility. Our scheme does not provide means to adjudicate between these options.   

Again, our definition allows us to recover the parts of the standard scheme that are physically well motivated: all symmetries associated with Noether's first theorem are, under our definition,  conservation symmetries.\footnote{Our scheme also allows us to capture classical symmetries exhibiting conserved charges that would not normally fall under the treatment of Noether's theorem. For example, General Relativity with asymptotically flat boundary conditions is locally invariant under spacetime diffeomorphisms. Nevertheless, it still has conserved charges associated to it (the ADM momenta \cite{adm:adm_review}) because the boundary variation is performed in a fixed way, using our terminology, due to how the asymptotic boundary conditions are imposed. Thus, the Poincar\'e invariance on the asymptotic boundary is a conservation symmetry by our definition.} An example particularly relevant to the considerations of this paper is that our classification scheme non-standardly directs us to categorise symmetries associated with temporal relabelling as conservation symmetries and, as well shall discuss later, this proves absolutely pivotal for understanding the role of time in \textit{relational} quantum theories, including prospective theories of quantum gravity.\footnote{The conserved quantity associated with relabelling symmetry is the Hamiltonian function itself, which, as one can easily show, is a conserved quantity of the classical evolution.} The key point is that, in general, conservation symmetries, since they are tied into the characteristic behaviour of the system, must be treated entirely differently from gauge symmetries when constructing the quantum formalism. They are associated with physical directions on phase space and should correspond to classical observables. In the quantum theory, there should, thus, be self-adjoint operators associated with the classical observables which act on the physical Hilbert space. One should therefore expect that general quantum states are formed by taking superpositions of the eigenstates of these operators. This is entirely unlike gauge symmetries, where the eigenvalues of the associated operators (i.e., the quantized generators of the classical symmetries) must be set to zero in the quantum formalism (following the Dirac quantization algorithm).

A simple case of a conservation symmetry will illustrate this point well. Let us consider a standard Newtonian point particle system treated as an isolated sub-system of the Universe consisting of three particles starting in different positions which evolve under the force of gravity. Now consider what happens if one performs time-dependent spatial translations to this system of particles, under which Newton's equations are manifestly invariant.

In our terms, such a translation is precisely a manifest symmetry since it corresponds to a global variation of configuration space curves in a direction along which the action is invariant. We can further classify the symmetry as manifest-fixed, because, for a Newtonian system, we are not free to vary the data on the end points -- the physics of the system places definite restrictions such that only some spatially translated variations are equivalent. In this case, these restrictions are just the conservation of linear momentum and the relevant constant of motion is just the total linear momentum of the system. Thus, we have a symmetry which is fixed and manifest -- a conservation symmetry in our terminology. Standard quantum mechanics, which is the quantum theory defined based upon Newtonian mechanics, is then such that we can have superpositions of momentum eigenstates, as one would expect from our general prescription. If, on the other hand, one has misclassified the spatial translations as gauge symmetries (i.e., manifest-free), then the quantum formalism that resulted would be a quantum theory of a single momentum eigenstate -- which is clearly not a faithful quantization of Newtonian theory. What is lost in this analysis is the ability to treat the centre-of-mass velocity of the system as an operator in the quantum theory, since forcing the system to a momentum eigenstate forces this observable to be precisely zero. However, if the three particle system is an isolated sub-system of the Universe, then the behaviour of the centre-of-mass velocity clearly has meaning as part of the characteristic behaviour of the system, and such a misclassification would fail to capture the full behaviour of the system. It is precisely this form of categorisation error that we hold to be behind the idea that `time disappears in quantum gravity', in complete analogy to how `centre-of-mass velocity disappears' in the example above.

The simplest case in our classification scheme is where there is no redundancy in the relevant representative relationship between a physical degree of freedom and its formal correlate. This is the case where there is no manifest symmetry and the variation is fixed. This case corresponds to a conventional dynamical degree of freedom. Its initial or boundary conditions are specified by the variational principle and, in a phase space formalism, it simply evolves according to the flow of the energy function or Hamiltonian. The different possibilities corresponding to the direction associated with the degree of freedom are physically distinguishable, and so must be counted as distinct  possibilities. Quantum mechanically, these degrees of freedom correspond to observable operators that can exist in the appropriate superpositions. 

The most non-trivial case is if there is no manifest symmetry and the variation is free. In this case, it is possible that there is a \emph{hidden} symmetry in the system. This can only happen if there is another manifest symmetry in the theory that has a particular type of formal relationship with the one at hand (it is \textit{second class} with respect to it -- i.e., the Poisson brackets of the constraints generating these symmetries is not weakly zero). If this is the case, the elements of the formalism can be modified (without changing the physical predictions of the theory) in such a way that the first symmetry becomes manifest. This is called \emph{symmetry trading} and has been used to construct the \textit{shape dynamics} formalism introduced in \cite{gryb:shape_dyn}. The general theory of symmetry trading is developed in \cite{Gomes:linking_paper}. The main formal result behind the symmetry trading formalism is that the two symmetries in questions can gauge fix each other, since they are generated by constraints that are second class with respect to each other. Thus, one has a \emph{choice} to interpret one of the constraints to be the generator of a symmetry while the other is the gauge fixing of it, but the opposite interpretive choice can equally be made. The two theories are equivalent because there is a special choice of gauge in both theories where the evolution on phase space is identical, given some initial data that solve the initial value constraints. We will give an more detailed description of this idea in the context of general relativity further down. For the time being, we can merely point out the clear interpretational implication of the existence of a hidden symmetry in terms of an \textit{underdetermination of symmetries}. One has the freedom to reformulate a theory such that different sets of transformations become symmetries. These new symmetries can be of either the gauge \textit{or} conservation type. The scheme does not provide means to adjudicate between these options.   

The quantum mechanical implications of trading hidden for non-hidden symmetries are subtle yet potentially very powerful. As shall be outlined below, one of the major possible benefits of symmetry trading is that it allows us to exchange one symmetry -- which we are unsure how to quantize, for another for which there are available techniques. Thus, the underdetermination of the symmetries of a theory at the classical level can give benefits regarding quantization: it gives one more options.  

Finally, if there is no manifest symmetry, the variation is free and there is no symmetry which can be traded, then the situation is more complicated because extra constraints need to be imposed in order for the symmetry to be preserved by the dynamics. These new constraints often lead, for the same reasons, to even further constraints leading to an infinite regress of constraints. This is a sign of an inconsistent system (i.e., a mathematically poorly posed formalism). We can now collect together all the possible types of degrees of freedom classified as distinct within our scheme in Figure 3. The appropriate interpretation implications and prescription for treating the quantum mechanical equivalents are then given in Figure 4.

\begin{figure}[h]
    \begin{center}
\begin{tabular}{|c|c|c|}\hline
  \textbf{Variation} & \textbf{Symmetry in Action} & \textbf{Classification}  \\\hline\hline
 Free & Manifest & Gauge Symmetry   \\\hline
 Fixed & Manifest & Conservation Symmetry \\\hline
 Fixed & None & Physical Direction \\\hline
 Free & Hidden & Tradable Symmetry \\\hline
 Free & None   & Possible Inconsistency\\\hline
    \end{tabular}
   
   \bigskip
   
   \small{Figure 3. New symmetry classification defined using physical principles.}\label{fig:table of dof}
    \end{center}
\end{figure}

\begin{figure}[h]
    \begin{center}
\begin{tabular}{|c|c|c|}\hline
  \textbf{Classification} & \textbf{Interpretational Implication} & \textbf{Quantum Degree of Freedom}  \\\hline\hline
 Gauge Symmetry  & Eliminate otiose variables  & Operator annihilates wavefunction   \\\hline
 Conservation Symmetry & Underdetermination of possibilities  & Operator obeying identity \\\hline
 Physical Direction & Fixed possibilities & Operator with no restriction \\\hline
 Tradable Symmetry  & Underdetermination of symmetries & Represented in new way?  \\\hline
 Possible Inconsistency & Formalism inconsistent?   & Quantization blocked   \\\hline
    \end{tabular}
   
   \bigskip
   
   \small{Figure 4. Implications of new scheme. Figure 4}\label{fig:table of implications}
    \end{center}
\end{figure}

The tables make clear both the generality and potential physical importance of our scheme. If a degree of freedom is misclassified then not only will the interpretation of its role within the classical formalism be incorrect, the quantum formalism derived will fail to capture the characteristic behaviour of the classical system in the appropriate limit. A mistake at this stage will lead to an incorrect quantum theory. The key claim that will be defended later in this paper is that precisely such a misclassification has been made for the case of gravity, and that the so-called timelessness of quantum gravity is actually a manifestation of this mistake, and not the absence of basic temporal structure in the relevant system class.

\subsection{Voluntary Redundancy} \label{sec: Voluntary Redundancy}

In the previous section, we outlined a classification scheme for the redundancies that can occur within the representative relationship between classical mechanical formalisms and classical physical systems. One of the most important applications for this scheme is to ensure a physically well-motivated quantization of the theory in question -- i.e., one leading to a quantum formalism where the relevant quantum mechanical analogues to the classical degrees of freedom are faithfully represented. Unfortunately, for some classical systems, standard techniques for quantization \textit{do not} lead to quantum formalisms with such properties.\footnote{This issue is over-and-above the occurrence of anomalies -- which we will not discuss here.} Thus, even if we correctly classify the symmetries in the classical formalism, we may lose track of them during the process of quantization. Furthermore, while, for simple systems and symmetries, it can be a straightforward task to isolate the degree of freedom associated with a particular symmetry, the general case is famously problematic when explicit systems are considered (as an example, compare electromagnetism to Yang--Mills theory). In order to prevent such problems, we recommend a general formal procedure that provides a concrete method for explicitly isolating the degree of freedom associated to the symmetry in question. The description we will give here for this procedure will be rather more intuitive than explicit. More technically inclined readers may refer to \S3 of the companion paper \cite{gryb:2014} for details.

The procedure we recommend for dealing with this problem involves introducing \emph{even more} redundancy into the formalism by introducing auxiliary fields that artificially parametrize the symmetry in question. These auxiliary fields, which are so named because they can be integrated out by inserting their classical equations of motion, are introduced into the theory in such a way that they shift all the degrees of freedom of the theory along the orbit of some gauge group in a way that we will describe in more detail below. Formally, these auxiliary fields `compensate' for the symmetry group in question, giving them a role very similar to that of a gauge \emph{compensator field}. These auxiliary degree of freedom can then either be varied in a \emph{free} way or \emph{fixed} way by either imposing or not imposing a specific functional restriction onto the corresponding compensator field. This restriction will be referred to as the \emph{best-matching constraint} because it originally appeared in the context of a procedure called \emph{best matching} developed by Barbour (for a nice introduction to best matching see \cite{Barbour:DefMach}). Although the use of these compensator fields is often a matter of finding a convenient way to mathematically isolate a degree of freedom associated with some symmetry, for the case of reparametrization symmetry -- which is the primary case of interest to us here -- the introduction of a compensator field is a \emph{mandatory} technical step in being able to faithfully represent the symmetry. 

To illustrate what the best-matching conditions achieve, we will briefly describe the role of the compensator fields in the formalism. These fields are simply symmetry group parameters that represent an active transformation of the configuration variables of the system. The group $\mathcal G$ itself is determined upon specification of one's formalism. In other words, once one chooses: i) the set of configuration variables one would like to use to describe a particular system and ii) a particular parametrization of the physical observables of the theory in terms of these variables, then the choice of group is given by the quotient of the former by the latter. In the case where $\mathcal G$ can be represented as a Lie group, the formalism becomes straightforward to describe. The active transformation of the configuration variables, $q$, can be written by exponentiating the contraction of the compensator fields, $\theta$, with the generators, $t\in\mathfrak g$, of the the relevant Lie algebra. 
This gives a set of actively transformed quantities $\bar q = e^{\theta\cdot\ t} q$ that depend on the compensator fields $\theta$. It is clear from this definition what these barred coordinates represent: the difference between them and the original $q$'s is just given by motion in the direction associated  with the symmetry group orbit (i.e., along the flow of the constraints generating the symmetry in question). Thus, they can be used to absorb the symmetric degree of freedom we are looking for.

For a simple example of how this works, consider the case of $N$ Newtonian particles in $1d$ with coordinates $q^i$ and suppose we want to consider the dilatations as a symmetry group. The action of this group on these coordinates is simply $q^i \to e^{\theta(t)} q^i$, where $\theta$ here is the compensator field parameterizing the dilatations. The effect this has on the theory is to perform time-dependent rescalings of the coordinates. We can ensure that this new field $\theta$ does not change the predictions of the theory (and is, thus, a genuine auxiliary field), by imposing an additional phase space constraint $\pi_\theta - \sum_i q^i = 0$ onto the theory, which is the infinitesimal generator of the gauge symmetry
\begin{align}
	q^i &\to e^{-\phi} q^i & \theta \to \theta + \phi
\end{align}
under which the quantity $\bar q^i = e^{\theta}q^i $ is clearly invariant. The net effect of this procedure is to mix up the part of the $q^i$ associated with the global scale with the new field $\theta$. The additional gauge symmetry above shows how this auxiliary field acts to compensate the (at this moment) artificial dilatational symmetry just introduced.


The compensator fields are then used to implement either a fixed or free variation in a two-step process. First, we define the momenta $p$, conjugate to $q$, and $\pi_\theta$, conjugate to $\theta$, and perform a canonical transformation  (i.e., a transformation that preserves the symplectic 2--form on the extended phase space) from the original coordinates $(q,\theta; p, \pi_\theta)$ to a set of barred coordinates $(\bar q, \bar \theta ; \bar p, \bar \pi_\theta)$. This canonical transformation is completely determined by the definition of $\bar q$ and the requirement that $\bar\theta = \theta$ (for details see \cite{gryb:2014}). The effect of this transformation is to effectively mix the compensator field with the symmetric degree of freedom in such a way that the barred momentum, $\bar \pi_\theta$, represents the momentum conjugate to the symmetric degree of freedom, as expressed in terms of the new variables. 

The second step is to perform the actual variation. How we do this depends crucially upon the free vs. fixed distinction. If the variation is fixed, we know that the degrees of freedom in the original phase space, $(q,p)$, were all physical. We should therefore treat only the directions associated with the introduction of the compensator fields (and their momenta) as corresponding to surplus representative structure. This equates to making the variation independent of two phase space degrees of freedom\footnote{For the remainder of the text, we will count phase space 
-- as opposed to configuration space -- degrees of freedom as is conventional in canonical approaches to quantum gravity.} for each compensator field, and can be done explicitly by enforcing the canonical restrictions $\pi_\theta=0$. This condition can be treated using standard gauge theory methods developed by Dirac \cite{Dirac:1964}. Although the details are not important here, the main result is that the restriction $\pi_\theta = 0$ eliminates the extra redundancy introduced into the theory when adding the compensator fields.\footnote{More explicitly, the condition $\pi_\theta = 0$ is a first class constraint which can be gauge-fixed by the condition $\theta = \bar\theta = 0$, thus eliminating two phase space degrees of freedom, as outlined in, for example, \cite{Henneaux:1992a}.} 

If the variation is free, then an additional step must be taken since the original theory already had non-physical redundancies. This step involves adding an additional constraint to the system that guarantees that the action is independent of the velocities of the $\bar\theta$'s. This will ensure that the theory is independent of any freely specifiable information associated with the symmetric degree of freedom, independently of when the end points of the variation are specified. Formally, we can express the relevant requirement as the disappearance of the transformed momentum variable to the compensator field, i.e., via imposing the \textit{best-matching} constraint equation: $\bar\pi_\theta = 0$.


Now, let us try and understand more clearly the role played by the compensator fields \textit{after} the transformation has mixed them with the symmetric degrees of freedom. The manifest symmetry requirement states that the action is invariant under the symmetry in question. In terms of the transformed compensator fields, this property, together with the Euler--Lagrange equations of the system, implies that $\bar\pi_\theta = \text{constant}$. This means that, in the manifest case, we always have a conserved quantity. This quantity is called the  \textit{Noether charge} associated to a global symmetry. 

In a fixed variation, this charge is determined by the initial conditions of the system. For example, if the symmetry in question is represented by linear translations in space, the momenta of the compensator fields correspond to the total linear momentum in each of the three spatial directions. Such quantities are, of course,  conserved in the evolution of any isolated system and we thus see that the conservation of the Noether charge relevant to linear translations simply is the expression of conservation of linear momentum. Classically, the value of a conserved charge is something definitely determined for once and for all time by the initial state of the relevant system. Quantum mechanically however, things are more flexible. Systems can, and generally do, exist in \textit{superpositions} of different values of the relevant charge. This is a direct manifestation of the fact that conservation symmetries are rooted in physical dynamics rather than redundant representative structure.

In the free case, however, there is no way to fix the corresponding charge. Instead, this is done by the best-matching conditions, which force the charge to vanish.\footnote{In electrodynamics, the best-matching constraint reduces to the usual Gauss constraint, so the terminology is a bit confusing: this `charge' does not correspond to the usual electrodynamic `charge'.} This has important implications for the quantum theory. The application of the quantum analogue of the best-matching conditions forbids the existence of superpositions of eigenstates of that charge, since only the zero eigenvalue is allowed. Below, we will see how this seemingly innocuous point becomes essential to understanding time and its denial in the context of quantum theories of gravity.

\section{Understanding Time}
\label{sec: Understanding Time}

\subsection{Change and Order} \label{sec: Change and Order}

Things change, and time -- whatever it may be -- certainly is, at a minimum, a means for describing this change. Newton was father to a notion of time that gives us much more than just a measure of change, since his absolute time `of itself, and from its own nature, flows equably without relation to anything external' \cite{Newton}. Such a Newtonian time can be taken to constitute an \textit{absolute temporal background} against which both the temporal order of, and the temporal distance between, events is defined irrespective of any changes that may take place. Thus, in a universe of no change, it makes sense to distinguish both the order of and time between events that are, other than their absolutely determined temporal position, entirely identical. More precisely, we can say that on the Newtonian view, both the \textit{metric} (distance) and \textit{topological} (ordering) structure of time are fixed absolutely irrespective of changes in the material universe.  

At the other end of the spectrum from a Newtonian notion of time, we can conceive of a radically relationalist conception along the lines discussed in Section 1. Recall that, on this correlation view of time, that is most closely associated with the work of Rovelli \cite{Rovelli:2002,Rovelli:2004}, all that it is for a physical degree of freedom to change is for it to vary with regard to a second physical degree of freedom -- and there is no sense in which this variation can be described in absolute, non-relative terms. Within this picture of the world, we explicitly deny \textit{both} time's  metric and topological structures. The notion of time one may recover from change is an inherently arbitrary and approximate one. We are free to choose any degree of freedom as an internal clock, and such an arbitrarily chosen clock may, for some finite interval, give us a useful means of marking both the duration between and order of events as defined by correlations between the other degrees of freedom. However, such an internal clock can only reliably give us an approximation of a temporal ordering, since it may always start to run backwards after some finite interval, and so generally will not give us an ordering of events in `time' that is globally defined in terms of a linear sequence: the parametrization is not \textit{monotonically increasing}. Moreover, such a method of distinguishing one variable as `the clock' inevitably involves neglecting the dynamics of precisely that variable: once the internal clock choice is made we are no longer able to describe the change in that clock degree of freedom. This is, in effect, to reduce the dimensionality of our system by one. Thus, in the context of this strong relationalism about time: a one-dimensional system is always static, a two dimensional system is \textit{really} a one dimensional system, and so on. 

Under a radical relationalist view, the ontology of time is constrained to be a very sparse one. Only relative variation, exists and there is no fundamental time ordering structure -- only ordering relative to an arbitrary and approximate internal clock. Adoption of such a view has direct implications for the `metaphysics of time' in that it is inconsistent with \textit{all three} of the notions in the famous McTaggart schema \cite{mctaggart:1908}. Recall that McTaggart distinguishes between, the dynamic time of the `A-series', the directed, but non-dynamic B-series, and the ordered but un-directed `C-series'. The C-series should reasonably be taken to be the minimal position since without such structure the richer metaphysics of time cannot be defined.\footnote{We are very grateful to Matt Farr for making us aware of this connection. See \cite{Farr:2012a,Farr:2012b,Farr:unpub} for insightful comments regarding McTaggart's C-series, its relation to the A and B Series, and details on the similar conception of time that appears in the work of  \cite{black:1959} and \cite{reichenbach:1956}.} However, the ordering structure required for the C-series is precisely what is denied within the radical relationalist view. Thus, we see that there is a precise sense that radical relationalism involves giving up time, rather than simply relativising it. If ordering structure (C-series change) is taken as an essential feature of the time concept, then since such a feature is inconsistent with radical relationalism, the view can be categorised as an essentially \textit{timeless} one.       

Let us now consider a third ontology of time. This viewpoint constitutes a `middle way', with less absolute time structure than the Newtonian, but more than the Radical Relationalist. Our starting point is set by the views of Ernst Mach. Mach both criticised absolute notions of time on epistemic grounds \textit{and} put forward a positive account of the kind of temporal structure we are presented with. According to Mach it is `utterly beyond our power to measure the changes of things by time...quite the contrary, time is an abstraction at which we arrive through the changes of things.' \cite{mach:mechanics}. Thus, on the Machian view, a consistent notion of time can be abstracted from change such that the inherently interconnected nature of every possible internal measure of time is accounted for. According to the Mittelstaedt--Barbour \cite{Mittelstaedt:machs_2nd,barbour:newton_2_mach} interpretation of Mach, we can understand this second Mach's principle as motivating a relational notion of time that is not merely internal but also equitable, in that it can be derived uniquely from the motions of the entire system taken together. Thus, any isolated system -- and, in fact, the universe as a whole -- should have its own natural clock emergent from the dynamics. The Machian view of time thus involves a relative notion of duration as abstracted from change. There is then a clear sense in which we can think of the Machian relationalist view of time as connected to the radical relationalist view: each involves a relational view of duration as derived from relative change. However, for a notion of time to be relational in the Machian sense, it is not enough to be merely relational -- it must also be unique and equitable. We cannot, therefore, merely identify an isolated subsystem as our relational clock, since to do so is not only non-unique but would also lead to an inequitable measure, insensitive to the dynamics of the clock system itself. This subtle difference regarding the Machian and radical notions of relational duration, encodes a more extreme difference regarding the status of time ordering. In assuming that there is a unique method for abstracting duration from change, the Machian view is also assuming that there is an absolute ordering within the change -- otherwise the abstraction process would be underdetermined. This means that the Machian relationalist view involves the assumption of absolute temporal ordering structure. The ordering of events is fixed absolutely irrespective of changes in the material universe. This is entirely different to the radical relationalist view where such structure is explicitly forbidden. Although the \textit{metric} structure of time is relativised, under the Machian view the \textit{topological} structure is presumed to remain absolute.   

We thus have three internally consistent views of time,  each containing a different level of absolute time structure: i) Newtonian (absolute duration, absolute ordering structure); ii) Machian relationalist (relative duration, absolute ordering structure); iii) radical relationalist (relative duration, relative ordering). Given a physical theory with time re-labelling symmetry, is there any restriction as to which of the three ontologies we can consistently interpret the theory in terms of? Clearly this depends upon the symmetry categorisation scheme which one adopts.      

According to one influential view\footnote{On this, see, in particular,\cite[\S3.2.4]{Rovelli:2004}.} for theories which feature temporal relabelling symmetries, radical relationalism is forced upon us by the mathematical structure of the theories in question. Since the time symmetries present in these theories are judged, according to the standard classification scheme, to be gauge symmetries, we are compelled to take the interpretational step of strong relationalism about time. As indicated above, to us the logic of such arguments seems a strange one. Reliance upon a purely mathematical prescription to determine our interpretation of the ontology associated with a physical formalism seems to put the cart before the horse. Moreover, it can actually be proved that the scheme in question \textit{does not} apply to at least the case of theories where global temporal relabelling is a symmetry \cite{Barbour:2008}.\footnote{This proof is an addition to independent arguments questioning the role of the Hamiltonian constraint as the generator of a gauge symmetry \cite{Pons:1997,PonsSalis:2005,Pons:2010,pitts:2014}. See \S5.1.} Our suggestion is that one should remain agnostic as to the `true nature of time' implied by a physical theory until after one has carefully considered the relationship between the formalism of that theory and the class of systems which it is taken to represent. In particular, one must have an understanding of the \textit{physical basis} of the relevant temporal symmetries in terms of how the characteristic behaviour of the system class are related to the abstract degrees of freedom, the boundary and initial (or end point) conditions, and the nomological restrictions on the evolution. Thus, on our view, it is only \textit{after} applying something like our physical prescription for classifying symmetries that one is able to consider the interpretation of the formalism in terms of a temporal ontology.

Let us do this explicitly for the case of a simple finite dimensional system where global temporal relabelling is a symmetry -- we will consider the more difficult case of infinite dimensional systems with local temporal relabelling symmetry, i.e., general relativity, in the following section. The situation is then this. We have a class of physical systems which have a finite number of physical degrees of freedom and for which there is no physical difference between the system passing through a sequence of physical states at different rates. Such a system may be represented via slightly adapted version of Newtonian mechanics called Jacobi's Theory.\footnote{This is after the great German mathematician Carl Jacobi. See \cite{Lanczos:1970} for details of both the formalism and its historical context. Jacobi's theory is called `relativistic mechanics' in \cite{Rovelli:2004}.} This theory can be constructed in terms of a configuration space with an identical number of degrees of freedom as a the corresponding Newtonian configuration space but with an action which displays an extra symmetry. This symmetry is \textit{reparametrization invariance}, and is equivalent to redefining the time parameter used to mark change between states within the theory.

The key question is then: what kind of symmetry is reparametrization invariance -- is it manifest? and is it free or fixed? The answer to the first question is fairly straightforward. Since, by construction, reparametrization invariance is a symmetry of the action, it can only be considered a manifest symmetry. What is interesting to note, however, is that the phase space direction associated with the symmetry is in fact given by the energy function or Hamiltonian. Thus the `symmetry direction' is also the `dynamics direction'. 

The next step is to determine whether we are dealing with a conservation or gauge symmetry by distinguishing whether we have a free or fixed variation. Again this is a fairly simple question to answer since, again by construction, we have a configuration space where all the abstract degrees of freedom directly correspond to physical degrees of freedom -- those of the corresponding Newtonian system. Explicitly, since infinitesimal change to the endpoints of the variational principle is in every direction parametrized by physical degrees of freedom, such variation cannot be done freely. It is fixed by the characteristic behaviour of the system, in particular the preparation conditions. Thus, reparametrization invariance, or temporal relabelling symmetry, is a conservation  and not a gauge symmetry.\footnote{ The conserved quantity associated with it varies in interpretation depending on the system in question. For timelike geodesics in spacetime, it corresponds to the mass of the test particle. For Jacobi mechanics, it corresponds to the energy of the system. In all cases, it is associated with the constrained Hamiltonian of the system.} This is precisely as one would expect since, as we have just noted, the phase space direction associated with the symmetry is precisely the direction of dynamical change within the theory -- thus, if the symmetry \textit{were} a gauge symmetry, then we should not expect to have any physical dynamics whatsoever since the `dynamics direction' would be an otiose representative structure. As we discussed above, in the case of a gauge symmetry we find identical physical possibilities being represented in terms of distinct instantaneous states. In this case, the supposedly identical physical possibilities are dynamically related states. In interpreting time relabelling symmetry as a gauge symmetry we would be interpreting dynamical change as unphysical. In that eventuality, the only option for an interpretation of the relevant temporal ontology \textit{would} be the radical relationalism mentioned above. However, by the lights of our scheme, such a move is formally unjustified. We \textit{do not} see temporal relabelling as a gauge symmetry and so can license an interpretation of theories with such a symmetry in terms of a more substantive notion of time. 

One important point regarding the conservation symmetry classification of reparametrization invariant theories relates to the interpretation of the role of energy. As noted above, all conservation symmetries have associated conserved quantities or charges. For the case in hand, the charge will be equivalent to the energy of the system. This means that the energy of the universe is interpreted as a constant of motion. This is in contrast to many existing views whereby, in a reparametrization invariant theory, energy is a constant of nature.  Classically, this difference has no empirical implications, but, quantum mechanically, it implies that we should expect superpositions of energy eigenstates -- as we would usually for conserved charges in the quantum version of a theory with conservation symmetry. We will return to this feature of what we call `relational quantum theories' in the following subsection.\footnote{One might wonder why one shouldn't expect the energy to simply be zero. At the moment, we have no physical principle to impose this condition. Indeed, in gravity, the analogue of the total energy (i.e., the homogeneous part of the Hamiltonian constraint) is the cosmological constant, which has been directly observed to be non-zero.}   

Before then, we must first consider the interpretational implications of our conservation symmetry approach to classical reparametrization invariant theory. Under our symmetry classification scheme, global reparametrization invariance is a conservation symmetry. The existence of global reparametrization invariance does not, therefore, motivate us to classify sets of instantaneous states as identical. Thus, the interpretation of a theory which displays global reparametrization invariance in terms of radical relationalism is not well motivated by our symmetry classification scheme. Such an interpretation would involve counting as identical possible instantaneous states which are not, by our scheme, related by gauge symmetry transformations. Thus, our symmetry classification scheme renders an interpretation of the given class of reparametrization invariant theories in terms of radical relationalism highly implausible.  

Our scheme does not, however, give us means to adjudicate between the two remaining options: as is typical for a conservation symmetry we have an \textit{underdetermination of possibilities}, and this underdetermination reflects precisely the difference between the Newtonian and Machian views. Under our scheme, physical theories which display global reparametrization invariance can be interpreted in terms of the Newtonian concept of time. Such an ontology of time involves counting as distinct possibilities \textit{histories} which are related by a conservation symmetry.\footnote{This is unless of course one where to adopt some sophisticated form of `temporal substantivalism' where, via the introduction of anti-Haecceitist reasoning about temporal points or otherwise, the possibility counting matches that of the Machian. See \cite{thebault:2012a,thebault:2012b}.} On the other hand, under our scheme, physical theories which display global reparametrization invariance can also be interpreted in terms of Machian relationalism. Such a view involves counting as identical \textit{histories} which are related by conservation symmetries. 

In the end, we are left with an interpretative choice regarding the status of temporal duration. To make this choice, further arguments in addition to the categorisation of symmetries are need. Here we will adopt the Machian view. This is because we find very plausible the Machian arguments in favour of an internal and equitable notions of time. Such arguments are essentially epistemological, and rely on the fact that the notion of duration we have available is always derived from change, and since there is no way for us to ever have access to an absolute duration measure, we would be better off doing without it. However, by relying on such motivations to do away with absolute duration, are we not opening the door to a charge of double standards. Is there not a parallel  epistemological worry regarding ordering structure: there seems to be no possible way for us to ever gain access to this `absolute time ordering' and so its adoption also seems to involve a rather strong metaphysical act of faith. There are two obvious ways of addressing this worry:

First, rather pragmatically, we can simply note that time ordering does appear to correspond to part of our physical formalism, and so, unless we can find an empirically adequate re-formulation without it, we have no good cause to question its status. There is, of course, a rather impressive precedent for the use of background structures to solve pressing theoretical problems, namely Newton's use of absolute space to give a coherent formulation of the principle of inertia \cite{rynasiewicz:1995a,rynasiewicz:1995b,pooleyfc}. And, in any case, if metaphysical minimalism is taken to be the main motivation for the elimination of non-empirical backgrounds, then accepting the fairly thin notion of a time ordering background is far more palatable than a full-strength Newtonian-style notion of time.   


More ambitiously, we could accept that time ordering should be founded in accessible features of our theory but that perhaps the arena for doing this is quantum and not classical theory. It is possible that the process by which temporal ordering in classical physics \textit{emerges} is connected to the classical limit of a fundamental quantum theory or even a broken symmetry in such a theory. A (rather technical) example that suggests the plausibility of this idea is in quantum field theory where a monotonic ordering is naturally encoded in the renormalization group (RG) flow near a conformal fixed point. Thus, the RG flow equation of a such a field theory could be reinterpreted as a time evolution equation in a shape dynamics theory. A simple toy model featuring such behaviour was studied in \cite{Barbour:2013qb}. Furthermore, there are exciting indications that a similar scenario could be used to reproduce certain models of inflation \cite{Skenderis:holo_uni}. However, many interesting open questions remain and such suggestions are still very tentative.

Given these two prospective justificatory strategies, let us accept, for the time being at least, that our middle-way, \textit{succession-as-absolute-and-duration-as-relative}, ontology of time can be coherently philosophically defended. What fruits can it bear when brought back into the domain of physical theory? Can it give us new insights into the nature of time in relational quantum theories?       

\subsection{Quantization and Succession}\label{sec: Quantization and Succession}

In the previous section, we defended both the classification of temporal relabelling symmetries as conservation symmetries and the interpretation of theories with such symmetries in terms of a temporal ontology within which duration is relative but succession absolute. Each of these moves gains significance for the future development of physical theory when seen in the context of quantization. This is particularly clear when considering the so-called problem of time that arises within attempts to quantize general relativity, but is also the case for simple globally reparametrization invariant models. The temporal relabelling symmetries of such models should, according to conventional classification schemes, be understood as gauge symmetries. This means that quantization, whether achieved via the voluntary redundancy route detailed above or otherwise, leads to a quantum theory in which the phase space directions defined by the flow of the Hamiltonian vector field (i.e. the null vector field associated with the Hamiltonian constraint) are treated as unphysical and, correspondingly, the quantum system is restricted to a zero eigenstate of the relevant charge. For the case of globally reparametrization invariant models, this is explicitly equivalent to treating dynamical directions as unphysical and to restricting the system to a single zero energy eigenstate. Thus, the classification of global temporal relabelling symmetries as gauge symmetries leads directly to a frozen quantum formalism. The only ontology we can associate to such a picture is that of radical relationalism -- and we are left without time. 

The alternative, for which we are arguing, is to treat global temporal relabelling as a conservation symmetry. The problem is then how to quantize the theory such that both the reparametrization symmetry and absolute temporal succession structure is retained. This is where the methodology of voluntary redundancy comes into its own. To our knowledge, the \textit{only} way to achieve quantization of a classical model such that time remains in the sense desired, is to use this method. Explicitly what we do (see companion paper \cite{gryb:2014} for more details) is choose a particular extension of the phase space such that our single configuration compensator field, $\tau$, has a canonical conjugate, $\pi_\tau$, proportional to the energy of the system. Our single constraint is then constructed by the combination of the original Hamiltonian plus the new momentum variable, so we have $H(q^i,p_i)+\pi_\tau=0$. As discussed in \S\ref{sec: Voluntary Redundancy} above, imposition of this constraint leads to a variational principle dependent upon all but two of the degrees of freedom -- i.e., precisely the number we started with. The quantum theory we reach by applying standard methods then preserves these physical degrees of freedom faithfully in that it allows their quantum analogues to change independently of each other. 

Furthermore, as expected, the quantum theory we arrive at is such that we can represent states in superpositions of eigenvalues of energy since total energy is the Noether charge associated with our conservation symmetry. Our quantum formalism can therefore accommodate fundamental temporal structure associated with succession via the parametrization chosen to distinguish the distinct energy eigenstates. However, since this parametrization is arbitrary, unlike in conventional quantum theory, there is no preferred classical temporal background which fixes a notion of duration. Time remains, but only in the form of succession. Because time is retained in a relational sense, we call this procedure for the quantization of theories with global reparamterization symmetries \textit{Relational Quantization}. From our perspective it is one of the crucial ingredients in the construction of a genuinely dynamic theory of quantum gravity as is consistent with the Machian view of time.                 

\section{Time and Gravitation}
\label{sec: Time and Gravitation}
\subsection{The Two Faces of Classical Gravity} \label{sec: Two Faces}

Our description of the universe is replete with different scales -- from the unimaginably small distances of particle physics, up to the unimaginably big distances involved in modern astronomy and cosmology. The theories relevant to these domains have one important feature in common: they treat such scales as an absolute background structure. Thus, in almost all modern physical theory, if we uniformly double the lengths involved, the phenomena will change.\footnote{More specifically, the Higgs and gravitational sectors are not conformally invariant.} Such scale dependence is also part of our best theory of gravity: general relativity. Although the theory incorporates a huge amount of descriptive freedom, it still privileges length scales. Interestingly however, the type of argument that drove Einstein to try and eliminate coordinate dependence from Newton's theory of gravity also motivates us to eliminate scale dependence from Einstein's. Just as he argued that we have no empirical access to absolute coordinate structures -- only relative ones; we may argue that we have no empirical access to absolute scale structures -- only relative ones.

There at least two distinct ways in which scale can enter a theory: the first is through the presence of dimensionful couplings, while the second is through the conformal factor of the metric (since this carries information about the absolute lengths of vectors). The former notion of scale invariance is intimately related to the renormalization group (RG) flow of a particular theory because it is the fixed points of the RG flow that are characterized by only dimensionless numbers. The latter is tied to geometry and whether or not lengths are preserved under parallel transport. In particular, invariance under local length scales can be expressed by requiring that gauge-invariant observables be invariant under the local symmetry
\begin{equation}\label{eq:Weyl}
	g_{ab}(x) \to e^{\phi(x)} g_{ab}(x),
\end{equation}
which is called a local \emph{Weyl} or often \emph{conformal} transformation. Although both notions of scale-invariance should be required in a truly scale-invariant theory, we will mostly be concerned with the more modest goal of implementing local Weyl invariance within our framework, leaving the more technically challenging problem of understanding the UV properties of our framework for the future.

What remains to be seen is to what extent local scale invariance can be implemented within a physically viable theory. Weyl's early attempts at a four dimensionally scale-invariant theory of gravity \cite{Weyl:1922} were ultimately unsuccessful, in terms of leading to a theory of quantum gravity, because of strong indications that the theory is unstable. Although work in this vein is ongoing \cite{Mannheim:2011ds,Hooft:2010nc} there are still significant issues to be overcome.

Here we will outline a proposal, different from Weyl's, which seeks to implement \textit{both} a Machian notion of time, and a three dimensional version of local scale invariance. The two ideas are in fact naturally connected.  As has been argued in \cite{Barbour:2011dn}, Mach's general stance of epistemic scepticism with regard to non-relational concepts, should lead us to the conclusion that it is the local scale invariant `shapes' of instantaneous configurations of the universe that should be taken as fundamental. These shapes can be determined by local observers through measurements of angles, which, on this view, are what is taken to be fundamental. The fundamental character attributed to angles can be justified through a simple epistemological observation: all measurements of lengths are, necessarily, local comparisons. Thus, real experiments only ever measure ratios of lengths in some local region, then use these local measurements to \emph{deduce} lengths for distant objects. Such arguments can be used to justify the expectation that real experiments \textit{should} be insensitive to local spatial Weyl transformations of the form \eqref{eq:Weyl}, which do not change the result of local measurements of ratios of lengths. What is left after removing the information about absolute local lengths are simply local angles.

In order for our definition of angles to be meaningful, there must exist a preferred notion of global time through which the instantaneous configurations can be defined. This is for the simple reason that spatial angles are clearly not invariant under local boosts or, more generally, under foliation-changing spacetime diffeomorphisms. Thus, we can see intuitively that the local temporal relabelling symmetry (which we understand to be explicitly foliation changing) is in conflict with three dimensional scale invariance. Just as observers using the spacetime picture of gravity must \emph{assume} the information about scale for distant objects, observers using the scale-invariant picture must \emph{assume} information about time for distance clocks by choosing a preferred time foliation.

Here we will examine this apparent conflict and offer conceptual foundations for its resolution (following the more formal arguments of \cite{gryb:shape_dyn}) by showing that these are really \emph{complimentary}, and not conflicting, pictures of reality. The first step in our reasoning relies upon the notion of hidden symmetry which has already been discussed briefly above. To recap, the idea is to identify, for a particular theory, a direction in which there is no manifest symmetry and the variation is free. In this case, it is possible that there is a hidden symmetry in the system. As discussed above, this can only happen if there is another manifest symmetry in the theory that has a particular type of formal relationship with the one at hand.\footnote{The formal requirement is that the constraint surfaces corresponding to the symmetries are `orthogonal', i.e., second class, on phase space. For the general theory behind symmetry trading, see \cite{Gomes:linking_paper}.} If this is the case, the elements of the formalism can be modified (without changing the physical predictions of the theory) in such a way that the first symmetry becomes manifest. Remarkably, it has been proved for the case of canonical gravity that, given certain reasonable simplicity assumptions, the unique set of hidden symmetries can be identified as three dimensional Weyl transformations which preserve the total volume of space (in the case of spatially closed topologies) \cite{Gomes:2013lka}. Intuitively, one can think of these as transformation that \emph{redistribute} scale from one region to another in a way that is very similar to what happens to the 2d surface of a balloon when the balloon is squashed or deformed. By definition, hidden symmetries can only be identified when there exists another symmetry which is manifest and has the required formal relationship such that the two sets can be understood as dual to each other. For the case in hand, the relevant dual to the scale symmetry is \textit{almost all} of the foliation symmetry. This means that if we \textit{symmetry trade} such that the hidden volume preserving scale transformation symmetry becomes manifest, we simultaneously switch to a theory with merely global, rather than local, time relabelling symmetry (this corresponds to a single, non-local Hamiltonian constraint). 

Specifically, what was proven in \cite{gryb:shape_dyn} is that there exits two theories on the phase space of General Relativity that are physically equivalent but have different symmetries: one is the standard ADM theory, which is foliation invariant, and the other is Shape Dynamics, which is invariant under (volume preserving) conformal transformations. The physical  equivalence of the formalisms is expressed by the fact that there is a special gauge choice in both theories where the dynamical trajectories on phase space are identical, given some valid initial data. It is possible, however, for these theories to differ if, for whatever reason, there are global obstructions to imposing the special gauges in both frameworks. These possibilities have been explored, for instance, in the case of black holes \cite{Gomes:2013fca}.

Once symmetry trading is completed, the theory we get has a neatly divided set of symmetries: volume preserving conformal transformations and spatial diffeomorphism symmetries can be classified as gauge symmetries; and global time relabelling can be classified as a conservation symmetry. This package of symmetries provides a clear constraint on the possible interpretations since it entails that spatial diffeomorphism and three dimensional conformal transformations connect physically identical instantaneous states, and leaves upon the interpretation of the global time relabelling symmetry.  This is in contrast to the symmetry package of general relativity. As discussed in \S1, there we have i) spatial diffeomorphism symmetry; and ii) local time relabelling symmetry. The first of these corresponds to a manifest free variation and thus, under our definition, is a gauge symmetry and therefore involves transformations between physically identical instantaneous states. This point is fairly uncontroversial (however see \cite{pitts:2013,pitts:2014}). Local time relabelling symmetry or re-foliation symmetry, on the other hand, in our scheme corresponds to a (very complicated) manifest fixed variation. This means that there is scope for a \textit{much} greater degree of underdetermination regarding the implications of the symmetry for possibility counting: the group of local time relabelling symmetries has an infinite number of parameters, the group of global relabelling symmetries only has one. Our categorisation is in contrast to the standard scheme where re-foliation is classified as a gauge symmetry, and thus, supposedly, gives us a means to identify physically identical instantaneous states. However, the sense in which re-foliations can be understood as connecting physically identical instantaneous states is a notoriously subtle and heavily qualified one \cite{thebault:2012b,pitts:2014}. Thus, the strength of our classification scheme in this instance is that it leaves open ambiguity regarding the implications of the  symmetry for possibility counting, exactly where it exists in practice. 

Let us recapitulate: there exists a unique formal move that allows us to re-describe gravitational systems in a fundamentally different way. In the language introduced earlier, we have an \textit{underdetermination of symmetries}. This first level of underdetermination is regarding a choice between two packages of symmetries. According to our understanding of symmetries there is then also an \textit{underdetermination of possibilities} -- depending upon how we interpret the conservation symmetries within the two packages. If we wish to break such underdetermination further motivations are needed.

We saw earlier that there a good Machian motivations for breaking the underdetermination of possibilities as related to global time-relabelling. Thus, as far as the shape dynamics package of symmetries is concerned we have a means to fix the interpretation of gravity fairly precisely. Let us consider a strategy for breaking the underdetermination of symmetries and see if we can motivate passage to the shape dynamics formalism. Above we argued that, in order to classify the symmetries relevant to a class of systems consistently, we must first  specify precisely what the relevant physical degrees of freedom and preparation  conditions are. This translates into asking what degrees of freedom are independently specifiable on the boundary of the variation or, in the case in hand, `What is fixed on the boundary in the action principles of General Relativity?' This question was posed by Wheeler to York, and is addressed in \cite{York:GR_boundary}. Our view corresponds to that taken in Section~4 of that paper: what is fixed on the boundary is:  i) a three geometry invariant under scale and coordinate labelling symmetries; and ii) the mean of the `York time' variable (which is canonical conjugate to the \textit{spatial} volume). We will refer to this identification of the independent degrees of freedom of gravity as \emph{York's ontology}. We take these variables to faithfully parametrize the characteristic behaviour of gravity and, thus, take their variation to be of the fixed kind, while variation with respect to all other variables is free.  York's identification of a locally scale-invariant three geometry as a variable to be fixed in the variational principle of gravity is consistent with the principle of scale-invariance just argued for on the basis of Mach's principles. However, one might note that York's second requirement: to keep the variable conjugate to the spatial volume fixed, is in direct conflict with the global principle of scale-invariance. Our view on this will be rather pragmatic at this stage. The fixing of this `York time' can be motivated by the directly observable red-shift, which is undeniably part of the characteristic behaviour of gravity. However, on Machian grounds, one might expect the red-shift to result as an emergent phenomenon from a fully scale-invariant theory. In this eventuality, we would still expect York's proposal to be valid in some effective limit in the quantum regime. However, since a concrete proposal where such a scenario is realized has not yet be developed, we will consider York's ontology directly.

From the York perspective, canonical general relativity has the rather undesirable feature of neither having manifest invariance under volume preserving conformal transformations, nor being a conventional gauge theory with respect to diffeomorphisms, nor varying the York time in fixed manner. The first two difficulties can be resolved by noting that the volume preserving scale transformations are a hidden symmetry of the canonical version of general relativity (as we have just noted). The last difficulty can be dealt with using our proposal for Relational Quantization procedure detailed in the previous section.

Before we embark on the final phase of our analysis and detail our specific proposal for time in a substantive sense, to remain within a theory of quantum gravity, let us take a moment to consider the philosophical consequences of symmetry trading on the way we should think of the ontology of classical gravity. The traditional understanding of general relativity (in canonical terms or otherwise) is as a theory of spacetimes invariant under spacetime  diffeomorphisms. This is essentially the ontology for gravity that was implied by Einstein's seminal work in the early part of last century, and which is still one key pillar of the scientific understanding of the universe. That this theory has a unique and robust formal correspondence, or duality, to another theory of gravity, which (under certain restrictions)\footnote{These restrictions amount to global foliability conditions of the spacetime by foliations where the trace of the extrinsic curvature of the hypersurfaces, as embedded into the spacetime, is constant.} has the same physical consequences is highly non-trivial. As mentioned in the introduction, on our view, it is taken to imply that gravity is essentially \textit{Janus-faced}. From this perspective, we should see the Einstein ontology of diffeomorphism invariant four dimensional spacetime geometries as only one face of gravity. The other, newly unveiled face being constituted by the ontology proposed by York: sequences of three dimensional spatial geometries accompanied by the specification of the York time variable and invariant under both diffeomorphism and scale transformations.

The situation of dual theories which are empirically equivalent, yet ontologically radically different is one of great interest within the philosophy of science since it seems to imply a particularly pernicious species of underdetermination.\footnote{See \cite{French:2011} for a excellent overview and analysis of various notions of `metaphysical underdetermination'  leading towards a motivation of the philosophical viewpoint of `ontological structural realism'. See \cite{Pooley:2006} and \cite[\S19]{thebault:2012thesis}  for discussion of the problems that the radically different ontologies found within theories of gravity may pose for the position.} Interestingly, the case which has garnered the most interest in the recent literature \cite{dawid:2007,rickles:2011,matsubara:2013}, that of the AdS/CFT correspondence in string theory/conformal field theory, is also one in which the two duals theories are respectively a diffeomorphsism invariant theory of gravity and a theory invariant under conformal transformations.\footnote{Indeed, it has recently been argued \cite{Gomes:2013qza} that the `bulk-bulk' equivalence of shape dynamics and general relativity can be used as an \emph{explanation} for certain limiting regimes of the AdS/CFT correspondence.} However, one would not have expected that such striking, and perhaps worrying, underdetermination scenarios could crop up in one of our most established physical theories. Yet, for our purposes, this seeming theoretical vice will prove a virtue. It is only by recognizing the second, scale-invariant face of gravity, that we can forge a new path towards quantization without sacrificing Time. 

\subsection{Retaining Succession in Quantum Gravity}\label{sec: Retaining Succession} 

As was argued for extensively in Section 2, correct identification and treatment of the physical degrees of freedom, and in particular their boundary variation behaviour, is essential to a faithful quantization of any physical theory. One of the major impediments to the quantization of gravity has been that, in canonical form, the symmetry and dynamics of the theory are `deeply entangled'.\footnote{Formally, this facet is encoded within the non-trivial structure of the Dirac--Bergmann constraint algebra, in particular that the bracket between two Hamiltonian constraints only closes with structure functions is indicative of the dual dynamics-symmetry aspect of these constraints. The constraint algebra of shape dynamics (and the relationally quantized theory) on the other hand is a genuine Lie algebra, and so a clear formal distinction between symmetry and dynamics can be made.} In the language of Section 2, this corresponds to the `fixed' and `free' aspects of the manifest symmetry being mixed together. Here we make no attempt to tackle the fearsome, and as yet unsolved, problem of a  disentangling the physical from gauge variations within the theory. Rather, given the `gravity is Janus-faced' revelation detailed above, we may simply turn to the alternative scale-invariant formalism for classical gravity and hope for a simpler representation of the relevant symmetries more amenable to quantization.

Starting with the canonical, or ADM, formulation of general relativity, symmetry trading yields a theory with the following manifest symmetries: a) three dimensional spatial coordinate invariance; b) three dimensional (spatial volume\footnote{For the case of open spatial topology, this global restriction turns into a specification of asymptotic boundary conditions \cite{Gomes:linking_paper}.} persevering) spatial conformal invariance; and c) invariance under one dimensional global time relabelling (or reparametrization). If we interpret this formalism in York's terms as discussed above, then we arrive at a unique and unambiguous specification of which of these symmetries correspond to free variations of the physical degrees of freedom at the boundary. By our symmetry classification scheme, such degrees of freedom are unphysical otiose variables. By construction, the York specification of boundary data is insensitive to variation of the (volume persevering) conformal and coordinate modes of the three dimensional metric tensor that characterises three dimensional spatial geometries. Thus, symmetries a) and b) are identified as gauge symmetries. Then we find, in correspondence with the discussion of Section 3, that the reparametrization symmetry c), is of the fixed kind. Thus, our prescription for quantization implies that we should introduce two global degrees of voluntary redundancy that parametrize the direction associated with the function that generates the reparametrization symmetry. As for the simple particle case, this function is a global Hamiltonian, and this means that our extra variables are a time-ordering label and a conserved charge associated with the `total energy'. We then append the second of these to the Hamiltonian to get an extended, but physically equivalent, formalism, which  then can be quantized (at least in formal terms) via standard methods. 

What we have gained in this rather circuitous route of symmetry trading, arbitrary extension and quantization, is simple to state. We are now equipped to represent the state of the universe \textit{at different times}. This is because, as for the particle case, we are able to take superpositions of energy eigenstates, and consider the independent evolution of observable operators. And yet, we \textit{have not} introduced a Newtonian-style background time into the theory. Time labellings are encoded in the arbitrary parameter which was introduced into the theory during the extension procedure, and so are neither fundamental nor observable.  Rather, the temporal structure which this formalism for quantum gravity contains is precisely the temporal succession structure we associated with the ideas of Mach above. As was noted in the opening section, we should not think of such a `temporal topological background' as being entirely alien to general relativity since it is in fact implicit within the canonical formalism at least (in terms of the positivity requirement on the lapse multiplier). Moreover, it is only by retaining such structure that, in the case of gravity, we can hope to preserve genuine change and avoid radical relationalism with regard to time, as we have been advocating. 

Our proposal for the quantization of gravity thus involves two  substantive interpretative moves. Firstly, the switch from the Einstein ontology (as implied by the general relativity formalism for gravity) to the York ontology (as implied by the shape dynamics formalism of gravity). Secondly, the promotion of time ordering (or topological) structure from an implicit formal feature to an explicit background. Both individually, and as a package, we can provide a range of motivations for these non-trivial steps. 

Most straightforwardly, there is the motivation from pragmatism: by following our prescription one opens up new strategies for theory development, and this, in the end, might be argued to be the true goal of foundational research. It remains to be seen precisely what lasting value the much vaunted `spirit of general covariance' will prove to have as a heuristic for \textit{future} theory construction. It may prove pivotal, or it may prove to have been misleading. Thus, if new and viable theoretical avenues can be opened up by reinterpreting symmetry in the context of gravity, we would be churlish to entirely ignore them since they do not sit conformably with an exciting abstract principle -- no matter how fundamental it may currently appear to be. We should not let fetishism for four-dimensional spacetime diffeomorphism invariance be a bar to potential progress. 

Furthermore, over and above the conceptual novelty of our proposal, it provides several notable formal advantages. In traditional approaches to the quantization of gravity (i.e. the `Wheeler-DeWitt-type' approaches), the resultant quantum formalism is such that only one energy eigenvalue is allowed. Evolution of the quantum states can then only be obtained by deparametrizing with respect to a degree of freedom, in the choice of which one must make an arbitrary decision. The definition of the functions used to represent observable quantities in the theory depend on this choice and, even for simple models, can lead to extremely complicated expressions.  Through our approach, we arrive at a formalism where there can be superpositions of energy eigenstates, and the evolution of the full state can be given with respect to the auxiliary time label. Thus, the evolution does not depend on any arbitrary choice of auxiliary time label. The identification of the relevant observables is then also non-arbitrary, and is technically much easier (because of the time-independence of the Hamiltonian). 

Finally, in addition to conceptual novelty and formal tractability, there is the simplest motivation of all: the motivation from time. It seems to be a basic requirement that, in one way or another, we are able to abstract some concept of time from our physical formalism -- without it our physics would simply fail to be descriptively adequate. If the only approach to the quantization of gravity were via a timeless formalism, then it would perhaps be fair to insist that we must make do with the  conceptual paucity of time merely as relative variation. However, given that there is a viable alternative route towards quantization, via symmetry exchange and relational quantization, the conceptual cost incurred by taking it should be counted as naught, next to the benefit of retaining a minimal, yet substantive, concept of time.

\section{Conclusion / Discussion}

\subsection{Related Arguments}

Here we would like to highlight related arguments appearing in recent work \cite{pitts:2014}, in addition to closely connected earlier observations \cite{Pons:1997,PonsSalis:2005,Pons:2010}. In certain key respects these authors reach  conclusions regarding time in general relativity that are closely related to our own. In particular, for them, as for us, the Hamiltonian constraint of GR is not understood to generate a gauge transformation, in the sense described by a local symmetry of the Lagrangian. The arguments of these authors focus upon the idea that, to get the Hamiltonian framework to match up with the Lagrangian notion of gauge symmetry, primary constraints must work in tandem with their associated secondary constraints to produce a genuine notion of gauge transformation. Although in many respects such arguments support essentially the same conclusions as those presented here, the chain of reasoning involved is largely independent. Contrasting and comparing the two approaches in a more detailed way would be an interesting topic for further study.

\subsection{Concluding Remarks}

The aim of this paper was to argue that there are strong formal and philosophical reasons to expect time to remain within any theory of quantum gravity. Although the temporal symmetries of  classical gravity are subtle, such that the redundant and physical aspects of the formalism are entangled, there does exist a precise formal recipe for making the unambiguous distinction needed for a faithful quantization. This recipe relies not just upon the technical  notions of symmetry trading and relational quantization discussed here (and presented more formally elsewhere), but also upon two quite general and simple philosophical morals. First, that physics is not mathematics: it is our understanding of how the physical formalism relates to the world that should govern our interpretation of its mathematical structures and not vice-versa. Second, that, at base, time has two aspects: metric and topological. While the first \textit{does} seem in conflict with the relational, and `background free' aspects of time in general relativity, the second appears implicitly even within the Einstein formalism. Furthermore, when seen in the context of our shape dynamics plus relational quantization proposal, topological or time ordering structure plays an important and unambiguous role: it is \textit{The Remains of Time} in quantum gravity. As such, the virtues of a program towards its conservation are self-evident.

\section*{Funding}

K.T. would like to thank the Alexander von Humboldt Foundation and the Munich Center for Mathematical Philosophy (Ludwig-Maximilians-Universit\"{a}t M\"{u}nchen) for support. S.G. would like to thank the The Netherlands Organisation for Scientific Research (NWO) and the National Sciences and Engineering Research Council of Canada (NSERC) for financial support and Utrecht University for accommodations.

\section*{Acknowledgements}

We are appreciative to Oliver Pooley and Hans Westman for valuable discussion of relevant points over the past years, to Seamus Bradley and Luca Moretti for comments on an earlier draft, and, in particular, to Julian Barbour for detailed critical comment. Detailed comments from two referees were also of great help. Thanks also to Eilwyn Lim for producing the Janus illustration above.

\bibliographystyle{chicago}
\bibliography{TimeRemains}

\begin{thebibliography}{}

\bibitem[\protect\citeauthoryear{Anderson}{Anderson}{2012}]{Anderson:PoTReview}
Anderson, E. (2012).
\newblock {`Problem of Time in Quantum Gravity'}.
\newblock {\em Annalen Phys.\/}~{\em \textbf{524}\/}(arXiv: 1206.2403), pp.
  757--86.

\bibitem[\protect\citeauthoryear{Anderson}{Anderson}{2013}]{anderson:2013}
Anderson, E. (2013).
\newblock {`Beables/Observables in Classical and Quantum Gravity'}.
\newblock {\em arXiv preprint arXiv:1312.6073\/}.

\bibitem[\protect\citeauthoryear{Anderson, Barbour, Foster, and
  O'Murchadha}{Anderson et~al.}{2003}]{barbour_el_al:scale_inv_gravity}
Anderson, E., J.~Barbour, B.~Foster, and N.~O'Murchadha (2003).
\newblock {`Scale-invariant gravity: Geometrodynamics'}.
\newblock {\em Class. Quant. Grav.\/}~{\em \textbf{20}}, p. 1571.

\bibitem[\protect\citeauthoryear{Anderson, Barbour, Foster, Kelleher, and
  O'Murchadha}{Anderson et~al.}{2005}]{barbour_el_al:physical_dof}
Anderson, E., J.~Barbour, B.~Z. Foster, B.~Kelleher, and N.~O'Murchadha (2005).
\newblock {`The physical gravitational degrees of freedom'}.
\newblock {\em Class. Quant. Grav.\/}~{\em \textbf{22}}, pp. 1795--802.

\bibitem[\protect\citeauthoryear{Arnowitt, Deser, and Misner}{Arnowitt
  et~al.}{1962}]{adm:adm_review}
Arnowitt, R.~L., S.~Deser, and C.~W. Misner (1962).
\newblock {`The dynamics of general relativity'}.
\newblock In {Louis Witten} (Ed.), {\em {Gravitation: an introduction to
  current research}}, pp.\  pp. 227--65. New York: John Wiley \& Sons.

\bibitem[\protect\citeauthoryear{Barbour}{Barbour}{2003}]{barbour:scale_inv_particles}
Barbour, J. (2003).
\newblock {`Scale-Invariant Gravity: Particle Dynamics'}.
\newblock {\em Class. Quant. Grav.\/}~{\em \textbf{20}}, pp. 1543--70.

\bibitem[\protect\citeauthoryear{Barbour}{Barbour}{2010}]{Barbour:DefMach}
Barbour, J. (2010).
\newblock {`The Definition of Mach's Principle'}.
\newblock {\em Found. Phys.\/}~{\em \textbf{40}}, pp. 1263--84.

\bibitem[\protect\citeauthoryear{Barbour}{Barbour}{2012}]{Barbour:2011dn}
Barbour, J. (2012).
\newblock {`Shape dynamics. An introduction'}.
\newblock In {\em Quantum Field Theory and Gravity}, pp.\  pp. 257--97. Basel:
  Springer.

\bibitem[\protect\citeauthoryear{{Barbour} and {Foster}}{{Barbour} and
  {Foster}}{2008}]{Barbour:2008}
{Barbour}, J. and B.~Z. {Foster} (2008, August).
\newblock {`Constraints and gauge transformations: Dirac's theorem is not
  always valid'}.
\newblock {\em ArXiv 0808.1223\/}.
\newblock ArXiv e-prints.

\bibitem[\protect\citeauthoryear{Barbour, Lostaglio, and Mercati}{Barbour
  et~al.}{2013}]{Barbour:2013qb}
Barbour, J., M.~Lostaglio, and F.~Mercati (2013).
\newblock {`Scale Anomaly as the Origin of Time'}.
\newblock {\em Gen.Rel.Grav.\/}~{\em \textbf{45}}, pp. 911--38.

\bibitem[\protect\citeauthoryear{{Barbour} and {Pfister}}{{Barbour} and
  {Pfister}}{1995}]{barbour:newton_2_mach}
{Barbour}, J.~B. and H.~{Pfister} (Eds.) (1995).
\newblock {\em {Mach's principle: From Newton's bucket to quantum gravity.
  Proceedings, Conference, Tuebingen, Germany, July 26-30, 1993}}.
\newblock Basel: Birkh{\"a}user.
\newblock Boston, USA: Birkhaeuser (1995) 536 p. (Einstein studies. 6).

\bibitem[\protect\citeauthoryear{Belot}{Belot}{2003}]{Belot:2003}
Belot, G. (2003).
\newblock {`Symmetry and gauge freedom'}.
\newblock {\em Studies In History and Philosophy of Modern Physics\/}~{\em
  \textbf{34}}, pp. 189--225.

\bibitem[\protect\citeauthoryear{Belot}{Belot}{2007}]{Belot:2007}
Belot, G. (2007, Jan).
\newblock {`The representation of time and change in mechanics'}.
\newblock In J.~Butterfield and J.~Earman (Eds.), {\em Handbook of Philosophy
  of Physics}, Chapter~2. Amsterdam: Elsevier.

\bibitem[\protect\citeauthoryear{Belot and Earman}{Belot and
  Earman}{2001}]{BelotEar:2001}
Belot, G. and J.~Earman (2001).
\newblock {`Pre-Socratic quantum gravity'}.
\newblock In C.~Callender and N.~Hugget (Eds.), {\em Physics Meets Philosophy
  at the Planck Scale}. {Cambridge: Cambridge University Press}.

\bibitem[\protect\citeauthoryear{Bergmann}{Bergmann}{1961}]{Bergmann:1961}
Bergmann, P.~G. (1961).
\newblock {`Observables in General Relativity'}.
\newblock {\em Rev. Mod. Phys.\/}~{\em \textbf{33}}, pp. 510--14.

\bibitem[\protect\citeauthoryear{Black}{Black}{1959}]{black:1959}
Black, M. (1959).
\newblock {`The ``direction'' of time'}.
\newblock {\em Analysis\/}~{\em \textbf{19}\/}(3), 54--63.

\bibitem[\protect\citeauthoryear{Bojowald, H{\"o}hn, and Tsobanjan}{Bojowald
  et~al.}{2011}]{bojowald:2011}
Bojowald, M., P.~A. H{\"o}hn, and A.~Tsobanjan (2011).
\newblock {`An effective approach to the problem of time'}.
\newblock {\em Classical and quantum gravity\/}~{\em \textbf{28}\/}(3), p.
  035006.

\bibitem[\protect\citeauthoryear{Dawid}{Dawid}{2007}]{dawid:2007}
Dawid, R. (2007).
\newblock {`Scientific Realism in the Age of String Theory'}.
\newblock {\em Physics and Philosophy\/}~{\em \textbf{501}}, p. 011.

\bibitem[\protect\citeauthoryear{DeWitt}{DeWitt}{1967}]{DeWitt:1967}
DeWitt, B. (1967).
\newblock {`Quantum theory of gravity. I. The canonical theory'}.
\newblock {\em Physical Review\/}~{\em \textbf{160}}, 1113--48.

\bibitem[\protect\citeauthoryear{Dirac}{Dirac}{1958}]{Dirac:1958b}
Dirac, P. A.~M. (1958).
\newblock {`The Theory of Gravitation in Hamiltonian Form'}.
\newblock {\em Proceedings of the Royal Society of London. Series A,
  Mathematical and Physical Sciences\/}~{\em \textbf{246}}, 333--43.

\bibitem[\protect\citeauthoryear{Dirac}{Dirac}{1964}]{Dirac:1964}
Dirac, P. A.~M. (1964).
\newblock {\em Lectures on quantum mechanics}.
\newblock New York: Dover.

\bibitem[\protect\citeauthoryear{Dittrich}{Dittrich}{2006}]{Dittrich:2006}
Dittrich, B. (2006).
\newblock {`Partial and complete observables for canonical general
  relativity'}.
\newblock {\em Classical and Quantum Gravity\/}~{\em \textbf{23}}, p. 6155.

\bibitem[\protect\citeauthoryear{Dittrich}{Dittrich}{2007}]{Dittrich:2007}
Dittrich, B. (2007).
\newblock {`Partial and complete observables for Hamiltonian constrained
  systems'}.
\newblock {\em General Relativity and Gravitation\/}~{\em \textbf{39}}, p.
  1891.

\bibitem[\protect\citeauthoryear{Earman}{Earman}{2003}]{Earman:2003}
Earman, J. (2003).
\newblock {`Tracking down gauge: an ode to the constrained Hamiltonian
  formalism'}.
\newblock In K.~Brading and E.~Castellani (Eds.), {\em Symmetries in physics},
  pp.\  pp. 150--62. {Cambridge: Cambridge University Press}.

\bibitem[\protect\citeauthoryear{Farr}{Farr}{2012a}]{Farr:2012a}
Farr, M. (2012a).
\newblock {`On A-and B-theoretic elements of branching spacetimes'}.
\newblock {\em Synthese\/}~{\em \textbf{188}\/}(1), 85--116.

\bibitem[\protect\citeauthoryear{Farr}{Farr}{2012b}]{Farr:2012b}
Farr, M. (2012b).
\newblock {\em Towards a C Theory of Time}.
\newblock Ph.\ D. thesis, University of Bristol.

\bibitem[\protect\citeauthoryear{Farr}{Farr}{UP}]{Farr:unpub}
Farr, M. (UP).
\newblock `the c theory of time' (unpublished).

\bibitem[\protect\citeauthoryear{French}{French}{2011}]{French:2011}
French, S. (2011).
\newblock {`Metaphysical underdetermination: why worry?'}.
\newblock {\em Synthese\/}~{\em \textbf{180}\/}(2), pp. 205--21.

\bibitem[\protect\citeauthoryear{{Geroch}}{{Geroch}}{1970}]{Geroch:1970}
{Geroch}, R. (1970, February).
\newblock {`Domain of Dependence'}.
\newblock {\em Journal of Mathematical Physics\/}~{\em \textbf{11}}, pp.
  437--49.

\bibitem[\protect\citeauthoryear{Giulini}{Giulini}{2013}]{Giulini:2013aa}
Giulini, D. (2013, 06).
\newblock {`Instants in physics - point mechanics and general relativity'}.
\newblock {\em arxiv.org/abs/1306.0338\/}.

\bibitem[\protect\citeauthoryear{Gomes}{Gomes}{2013}]{Gomes:2013lka}
Gomes, H. (2013).
\newblock {`A construction principle for ADM-type theories in maximal slicing
  gauge'}.
\newblock {\em arXiv 1307.1097\/}.

\bibitem[\protect\citeauthoryear{Gomes}{Gomes}{2014}]{Gomes:2013fca}
Gomes, H. (2014).
\newblock {`A Birkhoff theorem for Shape Dynamics'}.
\newblock {\em Class.Quant.Grav.\/}~{\em \textbf{31}}, p. 085008.

\bibitem[\protect\citeauthoryear{Gomes, Gryb, and Koslowski}{Gomes
  et~al.}{2011}]{gryb:shape_dyn}
Gomes, H., S.~Gryb, and T.~Koslowski (2011).
\newblock {`Einstein gravity as a 3D conformally invariant theory'}.
\newblock {\em Class. Quant. Grav.\/}~{\em \textbf{28}}, p. 045005.

\bibitem[\protect\citeauthoryear{Gomes, Gryb, Koslowski, Mercati, and
  Smolin}{Gomes et~al.}{2013}]{Gomes:2013qza}
Gomes, H., S.~Gryb, T.~Koslowski, F.~Mercati, and L.~Smolin (2013).
\newblock {`Why gravity codes the renormalization of conformal field
  theories'}.
\newblock {\em arXiv 1305.6315\/}.

\bibitem[\protect\citeauthoryear{Gomes and Koslowski}{Gomes and
  Koslowski}{2012}]{Gomes:linking_paper}
Gomes, H. and T.~Koslowski (2012).
\newblock {`The Link between General Relativity and Shape Dynamics'}.
\newblock {\em Class.Quant.Grav.\/}~{\em 29}, p. 075009.

\bibitem[\protect\citeauthoryear{Gryb and Th{\'e}bault}{Gryb and
  Th{\'e}bault}{2012}]{gryb:2012}
Gryb, S. and K.~Th{\'e}bault (2012).
\newblock {`The role of time in relational quantum theories'}.
\newblock {\em Foundations of physics\/}~{\em \textbf{42}\/}(9), pp. 1210--38.

\bibitem[\protect\citeauthoryear{Gryb and Th{\'e}bault}{Gryb and
  Th{\'e}bault}{2014}]{gryb:2014}
Gryb, S. and K.~Th{\'e}bault (2014).
\newblock {`Symmetry and Evolution in Quantum Gravity'}.
\newblock {\em Foundations of Physics\/}~{\em \textbf{44}\/}(3), pp. 305--48.

\bibitem[\protect\citeauthoryear{Henneaux and Teitelboim}{Henneaux and
  Teitelboim}{1992}]{Henneaux:1992a}
Henneaux, M. and C.~Teitelboim (1992).
\newblock {\em Quantization of gauge systems}.
\newblock Princeton: Princeton University Press.

\bibitem[\protect\citeauthoryear{Hooft}{Hooft}{2010}]{Hooft:2010nc}
Hooft, G.~t. (2010).
\newblock {`The Conformal Constraint in Canonical Quantum Gravity'}.

\bibitem[\protect\citeauthoryear{Isham}{Isham}{1993}]{Isham:pot_review}
Isham, C.~J. (1993).
\newblock {`Canonical quantum gravity and the problem of time'}.
\newblock In {\em Integrable systems, quantum groups, and quantum field
  theories}, pp.\  157--287. New York: Springer.

\bibitem[\protect\citeauthoryear{Lanczos}{Lanczos}{1970}]{Lanczos:1970}
Lanczos, C. (1970).
\newblock {\em The Variational Principles of Mechanics}.
\newblock New York: New York: Dover.

\bibitem[\protect\citeauthoryear{Mach}{Mach}{1883}]{mach:mechanics}
Mach, E. (1883).
\newblock {\em {Die Mechanik in ihrer Entwicklung Historisch-Kritsch
  Dargestellt}}.
\newblock Leipzig: Leipzig: Barth.
\newblock {English translation: Mach, E 1960 \emph{The Science of Mechanics},
  Open Court, Chicago (translation of 1912 German edition)}.

\bibitem[\protect\citeauthoryear{Mannheim}{Mannheim}{2012}]{Mannheim:2011ds}
Mannheim, P.~D. (2012).
\newblock {`Making the Case for Conformal Gravity'}.
\newblock {\em Found.Phys.\/}~{\em \textbf{42}}, 388--420.

\bibitem[\protect\citeauthoryear{Matsubara}{Matsubara}{2013}]{matsubara:2013}
Matsubara, K. (2013).
\newblock {`Realism, underdetermination and string theory dualities'}.
\newblock {\em Synthese\/}~{\em \textbf{190}\/}(3), 471--89.

\bibitem[\protect\citeauthoryear{McFadden and Skenderis}{McFadden and
  Skenderis}{2010}]{Skenderis:holo_uni}
McFadden, P. and K.~Skenderis (2010).
\newblock {`The Holographic Universe'}.
\newblock {\em J.Phys.Conf.Ser.\/}~{\em \textbf{222}}, p. 012007.

\bibitem[\protect\citeauthoryear{McTaggart}{McTaggart}{1908}]{mctaggart:1908}
McTaggart, J.~E. (1908).
\newblock The unreality of time.
\newblock {\em Mind\/}~{\em \textbf{17}}, 457--74.

\bibitem[\protect\citeauthoryear{Mittelstaedt}{Mittelstaedt}{1976}]{Mittelstaedt:machs_2nd}
Mittelstaedt, P. (1976).
\newblock {\em Der Zeitbegriff in der Physik}.
\newblock Germany: Mannheim: Wissenschaftsverlag.

\bibitem[\protect\citeauthoryear{Newton}{Newton}{1962}]{Newton}
Newton, I. (1962).
\newblock {\em {Sir Isaac Newton's Mathematical Principles of Natural
  Philosophy and his System of the World\emph{. Trans. Motte, A., revised
  Cajori, F.}}}
\newblock Berkeley: University of California Press.

\bibitem[\protect\citeauthoryear{Pitts}{Pitts}{2013}]{pitts:2013}
Pitts, J.~B. (2013).
\newblock {`A First Class Constraint Generates Not a Gauge Transformation, But
  a Bad Physical Change: The Case of Electromagnetism'}.
\newblock {\em arXiv preprint arXiv:1310.2756\/}.

\bibitem[\protect\citeauthoryear{Pitts}{Pitts}{2014}]{pitts:2014}
Pitts, J.~B. (2014).
\newblock {`Change in Hamiltonian General Relativity from the Lack of a
  Time-like Killing Vector Field'}.
\newblock {\em Studies In History and Philosophy of Modern Physics\/}~{\em
  \textbf{47}}, pp. 68--89.

\bibitem[\protect\citeauthoryear{Pons}{Pons}{2005}]{Pons:2005}
Pons, J. (2005).
\newblock {`On Dirac's incomplete analysis of gauge transformations'}.
\newblock {\em Studies in History and Philosophy of Modern Physics\/}~{\em
  \textbf{36}}, p. 491.

\bibitem[\protect\citeauthoryear{Pons and Salisbury}{Pons and
  Salisbury}{2005}]{PonsSalis:2005}
Pons, J. and D.~Salisbury (2005).
\newblock {`Issue of time in generally covariant theories and the
  Komar-Bergmann approach to observables in general relativity'}.
\newblock {\em Phys. Rev. D\/}~{\em \textbf{71}}, p. 124012.

\bibitem[\protect\citeauthoryear{Pons, Salisbury, and Shepley}{Pons
  et~al.}{1997}]{Pons:1997}
Pons, J., D.~Salisbury, and L.~Shepley (1997).
\newblock {`Gauge transformations in the Lagrangian and Hamiltonian formalisms
  of generally covariant theories'}.
\newblock {\em Phys. Rev. D\/}~{\em \textbf{55}\/}(2), 658--68.

\bibitem[\protect\citeauthoryear{Pons, Salisbury, and Sundermeyer}{Pons
  et~al.}{2010}]{Pons:2010}
Pons, J., D.~Salisbury, and K.~A. Sundermeyer (2010).
\newblock {`Observables in classical canonical gravity: folklore demystified'}.
\newblock {\em Journal of Physics A: Mathematical and General\/}~{\em
  \textbf{222}\/}(12018).

\bibitem[\protect\citeauthoryear{Pooley}{Pooley}{2006}]{Pooley:2006}
Pooley, O. (2006).
\newblock Points, particles and structural realism.
\newblock In S.~French, D.~Rickles, and J.~Saatsi (Eds.), {\em Structural
  Foundations of Quantum Gravity}, pp.\  pp. 83--120. Oxford: Oxford University
  Press.

\bibitem[\protect\citeauthoryear{Pooley}{Pooley}{FC}]{pooleyfc}
Pooley, O. (FC).
\newblock {\em The Reality of Spacetime (forthcoming)}.
\newblock Oxford: Oxford University Press.

\bibitem[\protect\citeauthoryear{Proust}{Proust}{1931}]{proust:1931}
Proust, M. (1931).
\newblock {\em Time regained}.
\newblock {L}ondon: {C}hatto {\&} {W}indus.

\bibitem[\protect\citeauthoryear{Reichenbach}{Reichenbach}{1956}]{reichenbach:1956}
Reichenbach, H. (1956).
\newblock {\em The Direction of Time}.
\newblock Berkeley: University of California Press.

\bibitem[\protect\citeauthoryear{Rickles}{Rickles}{2004}]{Rickles:2004}
Rickles, D. (2004).
\newblock {`Symmetry and Possibility: To Reduce or not Reduce?'}.
\newblock {\em philsci-archive.pitt.edu/archive/00001846/\/}.

\bibitem[\protect\citeauthoryear{Rickles}{Rickles}{2007}]{Rickles:2007}
Rickles, D. (2007).
\newblock {\em Symmetry, Structure, and Spacetime}.
\newblock Amsterdam: Elsevier.

\bibitem[\protect\citeauthoryear{Rickles}{Rickles}{2011}]{rickles:2011}
Rickles, D. (2011).
\newblock {`A philosopher looks at string dualities'}.
\newblock {\em Studies in History and Philosophy of Science Part B: Studies in
  History and Philosophy of Modern Physics\/}~{\em \textbf{42}\/}(1), pp.
  54--67.

\bibitem[\protect\citeauthoryear{Rovelli}{Rovelli}{2002}]{Rovelli:2002}
Rovelli, C. (2002).
\newblock {`Partial observables'}.
\newblock {\em Phys. Rev. D\/}~{\em \textbf{65}}, p. 124013.

\bibitem[\protect\citeauthoryear{Rovelli}{Rovelli}{2004}]{Rovelli:2004}
Rovelli, C. (2004).
\newblock {\em {Quantum Gravity}}.
\newblock {Cambridge: Cambridge University Press}.

\bibitem[\protect\citeauthoryear{Rovelli}{Rovelli}{2007}]{Rovelli:2007}
Rovelli, C. (2007).
\newblock {`Comment on ``Are the spectra of geometrical operators in Loop
  Quantum Gravity really discrete?'' by B. Dittrich and T. Thiemann'}.
\newblock {\em arxiv.org/abs/0708.2481\/}.

\bibitem[\protect\citeauthoryear{Rynasiewicz}{Rynasiewicz}{1995a}]{rynasiewicz:1995a}
Rynasiewicz, R. (1995a).
\newblock {`By their properties, causes and effects: Newton's scholium on time,
  space, place and motion---I. The text'}.
\newblock {\em Studies In History and Philosophy of Science Part A\/}~{\em
  \textbf{26}\/}(1), 133--53.

\bibitem[\protect\citeauthoryear{Rynasiewicz}{Rynasiewicz}{1995b}]{rynasiewicz:1995b}
Rynasiewicz, R. (1995b).
\newblock {`By their properties, causes and Effects: Newton's Scholium on time,
  space, place and motion---II. The context'}.
\newblock {\em Studies In History and Philosophy of Science Part A\/}~{\em
  \textbf{26}\/}(2), pp. 295--321.

\bibitem[\protect\citeauthoryear{Thebault}{Thebault}{2012}]{thebault:2012thesis}
Thebault, K.~P. (2012).
\newblock {\em Symmetry, Ontology and the Problem of Time: On the
  Interpretation and Quantisation of Canonical Gravity}.
\newblock Ph.\ D. thesis, University of Sydney.

\bibitem[\protect\citeauthoryear{Th{\'e}bault}{Th{\'e}bault}{2012a}]{thebault:2012a}
Th{\'e}bault, K.~P. (2012a).
\newblock {`Symplectic reduction and the problem of time in nonrelativistic
  mechanics'}.
\newblock {\em The British Journal for the Philosophy of Science\/}~{\em
  \textbf{63}\/}(4), 789--824.

\bibitem[\protect\citeauthoryear{Th{\'e}bault}{Th{\'e}bault}{2012b}]{thebault:2012b}
Th{\'e}bault, K.~P. (2012b).
\newblock {`Three denials of time in the interpretation of canonical gravity'}.
\newblock {\em Studies in History and Philosophy of Science Part B: Studies in
  History and Philosophy of Modern Physics\/}~{\em \textbf{43}\/}(4), pp.
  277--94.

\bibitem[\protect\citeauthoryear{Thiemann}{Thiemann}{2007}]{Thiemann:2007}
Thiemann, T. (2007).
\newblock {\em Modern canonical quantum general relativity}.
\newblock {Cambridge: Cambridge University Press}.

\bibitem[\protect\citeauthoryear{Weyl}{Weyl}{1922}]{Weyl:1922}
Weyl, H. (1922).
\newblock {\em Space-time-matter}.
\newblock New York, NY: New York: Dover.
\newblock Translation by Brose, Henry L from the original German text into
  English.

\bibitem[\protect\citeauthoryear{Wharton}{Wharton}{2009}]{Wharton:2009kp}
Wharton, K. (2009).
\newblock {`Extending Hamilton's principle to quantize classical fields'}.
\newblock {\em arXiv 0906.5409\/}.

\bibitem[\protect\citeauthoryear{York}{York}{1986}]{York:GR_boundary}
York, J. (1986).
\newblock {`Boundary Terms in the Action Principles of General Relativity'}.
\newblock {\em Found. Phys.\/}~{\em \textbf{16}\/}(3), pp. 249--57.

\end{thebibliography}

\end{document}